\documentclass[aps,amsfonts,pra,twocolumn,showpacs]{revtex4}
\usepackage{epsfig,amsmath,amssymb,bm,epsf,graphicx,psfrag}
\usepackage[all]{xy}
\usepackage{graphicx}
\usepackage{color}

\pdfoutput=1

\newtheorem{Theorem}{Theorem}

\newtheorem{Lemma}{Lemma} 

\newtheorem{Def}{Definition}

\newtheorem{Conj}{Conjecture}

\newtheorem{CP}{Computational primitive}

\pdfoutput=1

\begin{document}

\title{Symmetry-protected topological phases with uniform computational power in one dimension}

\author{$\text{Robert Raussendorf}^1$, $\text{Dongsheng Wang}^1$, $\text{Abhishodh Prakash}^{2}$, $\text{Tzu-Chieh Wei}^2$, $\text{David Stephen}^{1}$\\ \mbox{ }}

\affiliation{1: Department of Physics and Astronomy, University of British Columbia, Vancouver, BC, Canada,\\
2: C. N. Yang Institute for Theoretical Physics and Department of Physics and Astronomy, State University of New York at Stony Brook, Stony Brook, NY 11794-3840, USA}

\date{\today}

\begin{abstract}We investigate the usefulness of ground states of quantum spin chains with symmetry-protected topological order (SPTO) for measurement-based quantum computation. We show that, in spatial dimension one, if an SPTO phase supports quantum wire, then, subject to an additional symmetry condition that is satisfied in all cases so far investigated, it can also be used for quantum computation.
\end{abstract}


\maketitle

\section{Introduction}

The computational power of measurement-based quantum computation (MBQC) \cite{RB01} critically depends on the resource state used. Families of resource states enabling universal quantum computation exist---cluster states \cite{BR} and AKLT-states \cite{AKLT} (see \cite{Aka0}, \cite{WAR}) are examples---but they are very rare in Hilbert space \cite{GFE}, \cite{BMW}.

The latter changes in the presence of symmetry. Here, we are concerned with the scenario of symmetry-protected topological order (SPTO) \cite{SPT1}-\cite{SPT3}, in spatial dimension one. In this setting, Hamiltonians are constrained to be invariant under a group $G$ of symmetry transformations, and we consider the ground states of such Hamiltonians as resource states for MBQC. These ground states form physical phases, across which their properties  vary smoothly.

It has been conjectured \cite{Aka1}, \cite{Else} that in the scenario of sym\-metry-protected topologic order, physical phases are identical with ``computational phases''. Namely, 
\begin{Conj}\label{MC} 
The computational power of ground states for measurement-based quantum computation is uniform across each symmetry-protected topologically ordered phase.
\end{Conj}
The evidence for this conjecture is the following. In spatial dimension one, for any symmetry-protected phase characterized by a finite abelian symmetry group and a maximally non-commutative cohomology class, all matrix-product states in the phase support quantum wire \cite{Else}. That is, quantum information can be shuttled from one end of the spin chain to the other by local measurements, and this process is robust against symmetry-respecting perturbations of the ground state. With regard to quantum computation, it has been shown that for one specific symmetry group, $S_4$, there exists a computationally universal SPT ordered phase \cite{Aka3}. Also, there is an extended universal region for a particular $A_4$ invariant model \cite{PraWei} (for a phase diagram of SPT-ordered phases in one dimension under the symmetry group $A_4$ see \cite{PraWei2}). 

There is also support for Conjecture~\ref{MC} in spatial dimension ${\cal{D}}= 2$. It is known that the two-dimensional AKLT state \cite{AKLT} is a universal resource for MBQC, on various lattices \cite{Aka0}, \cite{WAR}, \cite{WR2}, \cite{WR3}. Now, the AKLT state can be deformed into a one-parameter family of quantum states such that for a sufficiently strong deformation in one direction, the state transitions from the valence-bond phase into the N{\'e}el-ordered phase \cite{Zitartz}.  It has been numerically demonstrated that along this line of deformed states, the transition of computational power coincides with the physical phase transition \cite{Darmawan}, \cite{HWW}.

These findings have led to the notion of ``computational phases of quantum matter'', which represents an intriguing connection between condensed matter physics and quantum computation.

Here, we further corroborate the above conjecture through the following result. In spatial dimension one, if an SPTO phase supports quantum wire, then, subject to an additional symmetry condition which is satisfied in all cases so far investigated, it can also be used for quantum computation. This result, given as Theorem~\ref{MT} in Section~\ref{SR}, is a strengthening of both \cite{Else} and \cite{Aka3}. Namely, it promotes the result  \cite{Else} from quantum wire to quantum computation. Further, it shows that computational universality in one-dimensional symmetry-protected topologically ordered systems does not only occur for one particular symmetry group \cite{Aka3}, but rather is an ubiquitous phenomenon. 

The main technical contribution of this paper is a method for carrying out MBQCs suited to resource states in SPT ordered phases, namely the incoherent addition of computational paths. It leads to a number of computational primitives, the first of which is the ``oblivious wire'' described in Section~\ref{FW}. Oblivious wire  is the basis for performing unitary gates (Section~\ref{iG}) and projective measurements (Section~\ref{M}) on the virtual space of the MPS description. It is a counterpart of ``buffering'' described in \cite{Aka3}. Its main advantage is that it can be applied in every SPTO phase that has a basis for quantum wire. 

It should be noted at this point that an MBQC resource state laid out in spatial dimension ${\cal{D}}$ leads to the simulation of a quantum circuit in spatial dimension ${\cal{D}}-1$, and hence in ${\cal{D}}=1$ only a single logical qudit is evolved. The cases of dimension ${\cal{D}} \geq 2$ are therefore of greater practical interest. However, numerous technical aspects are easier to handle in dimension one, and this case can therefore serve as a testing ground for novel computational techniques.

Besides corroborating the notion of ``computational phases of matter'', there is a second motivation for the present work. It is the prospect of classifying MBQC schemes by symmetry. 

If Conjecture~\ref{MC} is true,  not only are universal resource states easy to come by in the symmetry-protected case, but also,  computational power becomes a property of a whole quantum phase as opposed to individual quantum states. Moreover, quantum phases have a succinct mathematical characterization in SPTO, namely, in ${\cal{D}}=1$, every such phase is uniquely specified by the symmetry group $G$ and an element in the second cohomology group $H^2(G,U(1))$ of $G$.  If Conjecture~\ref{MC} holds, the computational power of the corresponding MBQC scheme is characterized by the same mathematical objects, giving rise to a classification of MBQC schemes by symmetry.

\section{Statement of the problem and the result}\label{SR} 

\subsection{Starting point and problem}

We consider symmetry protected topological phases {with a unique ground state}, in one spatial dimension. Be $|\Phi\rangle$ a ground state state in such a phase, given in matrix product form
\begin{equation}\label{Phi}
|\Phi\rangle =  \kappa(n) \sum_{i_1,\, ..\,, i_n}  \langle R| A[i_n]\cdot .. \cdot A[i_1]|L\rangle\,  |i_1,..,i_n\rangle,
\end{equation}
where $\kappa(n)$ is a normalization constant, and the $A[i]$, $i=1,\,..\,, d$ are $D_b\times D_b$ matrices. Therein, $d$ is the Hilbert space dimension of the local physical system, and $D_b$ is the bond dimension, i.e., the dimension of the virtual system. 

{Uniqueness of the ground state requires that the corresponding MPS tensors $A[i]$ satisfy the injectivity condition \cite{MPS}. That is, possibly after blocking $K$ consecutive sites, the MPS tensors $A[\textbf{i}]:=\prod_{k=1}^K A[i_k]$ span the space of $D_b\times D_b$ matrices.}

Measurement-based quantum computation may be run on the states Eq.~(\ref{Phi}) as usual; i.e., the simulated gates are chosen by the local measurement bases, taking into account previous measurement outcomes.

Among all the local bases in which the physical degrees of freedom can be measured, there is, for suitable SPTO phases, a special one, namely the so-called wire basis. \begin{Def}\label{WireBasis}
A wire basis is an orthonormal basis (ONB) ${\cal{B}}=\{|0\rangle,\,..\,,|d-1\rangle\}$ of the physical system such that the matrices $A[i]$ factorize into a logical part $C_i$ and a junk part $B_i$,
\begin{equation}\label{VSdec}
A[i] = C_i \otimes B_i,
\end{equation} 
where the $C_i$ are all unitary and constant across the phase. 
\end{Def}
The significance of the wire basis is that it leads to wire protected by the symmetry throughout the SPTO phase. Specifically, the following result has been established.
\begin{Theorem} {\em{\cite{Else}}}. \label{WT}
Consider a symmetry-protected phase characterized by a finite Abelian group and a maximally non-commutative cohomology class $[\omega]$. Then, for every MPS in this phase there exists a wire basis w.r.t. which the MPS tensor $A$ has the decomposition Eq.~(\ref{VSdec}). The unitary byproduct operators $C_i$ therein are elements of a finite group.
\end{Theorem}
Measurement of the physical degrees of freedom in the wire basis thus implements wire on the logical part of the virtual system. By virtue of Eq.~(\ref{VSdec}), the tensors $A[i]$ never entangle the logical subsystem 
with the uncontrolled junk subsystem, and thus preserve the logical information.\medskip

Now we turn to the problem: Since the byproduct operators $C_i$ generate a finite group, the above construction does not achieve universal quantum computation on the logical subsystem. There is no immediate fix for this. For example, if the wire basis is perturbed to implement a continuous set of operations, with respect to the new bases $\{|i'\rangle\}$ the tensors $A$ no longer satisfy the factorization property Eq.~(\ref{VSdec}), $A[i']\neq C_{i'} \otimes B_{i'}$. In result, the logical subsystem becomes entangled with the junk subsystem under measurement in such bases, and quantum information is lost. Thus, the following question arises: ``{\em{Can quantum wire \cite{Else} for the logical subsystem be promoted to universal quantum computation?''}}. 

The present paper gives an affirmative answer to this question. We point out that the desired extension from wire to computation is already known for one specific SPTO phase of one symmetry group, $S_4$ \cite{Aka3}. The present solution is widely applicable. It only assumes the existence of a wire basis in the considered SPTO phase, and a symmetry condition to be explained below.

\subsection{Result}

To state our result, we need to make two more definitions. We first capture the notion of ``uniformity of computational power'' of a physical phase.
\begin{Def}
A given physical phase has uniform computational power $X$ with respect to MBQC if all states in this phase, with the possible exception of a set of measure zero, have computational power $X$.
\end{Def}

The finite set of logical byproduct operators $C_i$ appearing in the decomposition Eq.~(\ref{VSdec}) will be used to realize a continuous set of quantum gates in MBQC, acting on the logical part of the virtual space.
\begin{Def}\label{Lie}
${\cal{O}}$ is the set of Hermitian operators
\begin{equation}\label{Odef}
\frac{C_i^{-1}C_j+C_j^{-1}C_i}{2}, \frac{C_i^{-1}C_j-C_j^{-1}C_i}{2i},
\end{equation}
 for all $0 \leq i <j \leq d-1$.
${\cal{A}}({\cal{O}})$ the algebra generated by the operators in ${\cal{O}}$ under $[\cdot,\cdot]/i$ and linear combination with real coefficients, and $L({\cal{O}})=\exp(i{\cal{A}}({\cal{O}}))$ is the Lie group generated by ${\cal{A}}({\cal{O}})$.
\end{Def}
Returning to the symmetric MPS state $|\Phi\rangle$ described in Eq.~(\ref{Phi}), if  injectivity holds then  the tensor $A$ has the symmetry
\begin{equation}\label{SymmA}
V(g)^\dagger A[|\psi\rangle]V(g) = A[u(g)|\psi\rangle],\;\; \forall g\in G,
\end{equation}
where $V(g)$ is some projective representation of the symmetry group $G$ acting on the virtual system and $u(g)$ is a unitary representation of $G$ acting on the physical degrees of freedom \cite{SPT1}, \cite{SPT3}.

With these definitions and observations, we can now state our main result.
\begin{Theorem}\label{MT}
Consider a symmetry-protected phase of a group $G$ with the properties (i) the ground state is unique, (ii) there is a wire basis, and (iii) for all $i=0,..,d-1$ exists a $g\in G$ such that $C_i\otimes I=V(g)$. Then, this SPTO phase has the uniform computational power to execute MBQC simulations of the unitary gates $L({\cal{O}})$ and the projective measurements of all observables in ${\cal{O}}$, with arbitrary accuracy.
\end{Theorem}

{\em{Remark:}} The symmetry condition, $C_i \otimes I \in V(G)$, for all $i = 0,..,d-1$, ensures that the randomness of measurement can be accounted for in MBQC in the standard fashion \cite{RB01}. However, there exist other ways of accounting for this randomness in MBQC \cite{Leung},\cite{Eisert}, and the symmetry condition in Theorem~\ref{MT} may therefore be unnecessarily stringent. On the other hand, it is satisfied in all SPTO phases with a wire basis so far encountered.\smallskip

In a companion paper \cite{QCcoho}, we describe how the MBQC-simulable Lie group $L({\cal{O}})$ of unitary gates is determined by the characteristics of the SPTO phase in which the resource state lives.

\section{Strategy for turning wire into computation}

\subsection{Computational primitives}

{The extension from quantum wire to logical gates proceeds in several steps. The key technical ingredient is the ``incoherent addition of computational paths''. A computational path is simply the measurement record $\textbf{s}$ obtained in an individual run of the given MBQC. ``Addition of paths'' means that, after correcting/ accounting for the outcome-dependent byproduct operator (which always arises in MBQC), the measurement record is discarded, and multiple computational paths are combined. For interpretation of this procedure, see Section~\ref{F/D}.}

{The first application of adding computational paths is the ``oblivious wire'' described in Section~\ref{FW}. It provides the capacity to drive any state of the virtual system into a tensor product state $\sigma\otimes \rho_\text{fix}$ between the logical and the junk subsystem. In particular,  the junk system ends up in the same fixed point state $\rho_\text{fix}$ every time the oblivious wire is applied. The oblivious wire thus provides a means of conditioning the junk subsystem, which turns out to be of importance for implementing quantum gates. In the present construction, oblivious wire is the counterpart to the buffering technique employed in \cite{Aka3}. In contrast to buffering, oblivious wire requires no trial-until-success.}

{The oblivious wire leads to three computational primitives, namely (i) the preparation of the virtual system in a tensor product state between the logical and the junk subsystem, (ii) unitary gates on the logical subsystem, and (iii) measurement of the logical subsystem. The latter can also be used for initialization. These three computational primitives are described in Sections~\ref{FW}, \ref{iG} and \ref{M}, respectively.} 

\subsection{Boundary conditions}

\begin{figure}
\begin{center}
\includegraphics[width=8.4cm]{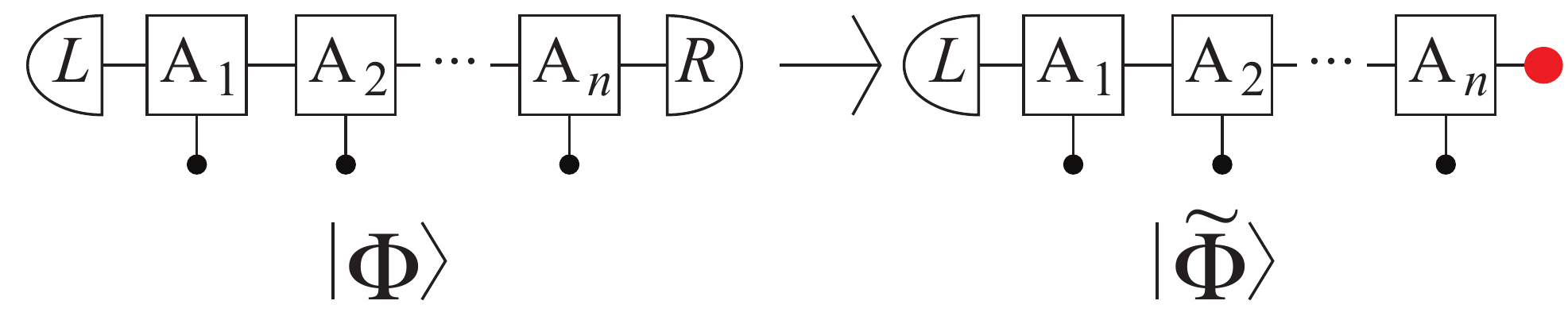}
\caption{\label{Bdy}Boundary conditions of the resource states $|\Phi\rangle$ and $|\tilde{\Phi}\rangle$. $|\tilde{\Phi}\rangle$ has an additional degree of freedom on its right boundary whose number of states is equal to the bond dimension $D_b$.}
\end{center}
\end{figure}

{The oblivious wire, the computational primitives, and the composition of these primitives are intertwined matters. In order to discuss them one after the other rather than everything all at once, we apply a technical trick. Namely, we temporarily change the matrix-product representation of the resource state at the right boundary; See Fig.~\ref{Bdy}. Specifically, we add a physical degree of freedom at the right boundary, whose dimension equals the dimension of the virtual system. The resulting state $|\tilde{\Phi}\rangle$ has the matrix product representation
\begin{equation}\label{PhiTil}
|\tilde{\Phi}\rangle =  \kappa(n) \sum_{i_1,.., i_n}  A[i_n]\cdot .. \cdot A[i_1]|L\rangle \otimes |i_1,..,i_n\rangle.
\end{equation}}
{Physically, this means that MBQC on $|\tilde{\Phi}\rangle$ can be run as a state preparator, in which the output state is mapped from the virtual system to the physical system at the right boundary. From a practical perspective, the need for a physical system of potentially high dimension on the right boundary is unappealing, since it places additional and unnecessary requirements on the experimental setup.}

However, we emphasize that the change of boundary conditions from Eq.~(\ref{Phi}) to Eq.~(\ref{PhiTil}) is employed here only as a tool for reasoning, and it is temporary. 
We revert to the original boundary conditions of the resource state $|\Phi\rangle$ of Eq.~(\ref{Phi}) in the very last step of the argument; See Section~\ref{Cc}. Our main result, Theorem~\ref{MT}, applies to the standard boundary conditions of Eq.~(\ref{Phi}).

\subsection{Oblivious wire}\label{FW}

The first computational tool to be established is the ``oblivious wire'', which prepares the virtual system in a tensor product state between logical and junk subsystem, and drives  the junk system towards a fixed point state. Below, we discuss three procedures of implementing wire, which are minor modifications of another. The first is the MBQC wire proper, and the third is the ``oblivious wire'', the procedure of our interest.

{\em{Procedure I:}} All measurement outcomes are fully remembered. All spins are measured in the wire basis, with outcomes $s_1,\,..\,,s_n$. The only unmeasured system then is the right boundary, and it is in the state
$$
|R\rangle =\Sigma(\textbf{s}) \otimes \left(\prod_{k=1}^n B_{s_k}\right) |L\rangle.
$$
Therein, the first tensor factor is for the logical and the second tensor factor for the junk subsystem. $\Sigma(\textbf{s})$ is the cumulative byproduct operator on the logical system, depending on the set $\textbf{s}$ of measurement outcomes,
\begin{equation}\label{Sigma}
\Sigma(\textbf{s}) = \prod_{k=1}^n C_{s_k}.
\end{equation}
So, the action on the logical system is that of the identity modulo a byproduct operator, all across the given physical phase.

{\em{Procedure II:}} The spins 1 to $n$ are measured as before. But then, in addition, the outcome-dependent byproduct operator $\Sigma(\textbf{s})$ is reversed on the right boundary system by applying its inverse. That is, the total action on the state $|\tilde{\Phi}\rangle$ is $|\textbf{s}\rangle \langle \textbf{s}|\otimes \Sigma(\textbf{s})^{-1}$, where $|\textbf{s}\rangle := \bigotimes_i |s_i\rangle$. This evolution puts the right boundary system in the state
$$
|R\rangle = I \otimes \left(\prod_{k=1}^n B_{s_k}\right) |L\rangle,
$$
for all sets $\textbf{s}$ of measurement outcomes. This is now an exact wire on the logical system.

{\em{Procedure III:}} Procedure II is run, and then outcomes $\textbf{s}$ are subsequently discarded.\medskip 

What is being implemented on the state $|\tilde{\Phi}\rangle$ in Procedure III is the quantum channel with Kraus operators
\begin{equation}\label{Alice}
P_\textbf{s} = |\textbf{s}\rangle \langle \textbf{s}| \otimes \Sigma(\textbf{s})^{-1}.
\end{equation}
Here, the first tensor factor is for all the physical spins in the chain, and the second tensor factor is for the additional physical degree of freedom located on the right boundary. 
It is easily verified that $\sum_\textbf{s} P_{\textbf{s}}^\dagger P_{\textbf{s}} = I$, as required for a POVM. The resulting quantum state of the right boundary system is 
\begin{equation}\label{taur}
\begin{array}{rcl}
\tau_R &=& \displaystyle{\text{Tr}_{1,\,..\,,n} \sum_{\textbf{s}} \left(|\textbf{s}\rangle \langle\textbf{s}| \otimes \Sigma(\textbf{s})^{-1}\right) |\tilde{\Phi}\rangle \langle \tilde{\Phi}| \left(|\textbf{s}\rangle \langle\textbf{s}| \otimes\Sigma(\textbf{s})\right)}\\
&=& \displaystyle{\sum_{\textbf{s}} \Sigma(\textbf{s})^{-1}  \langle\textbf{s}|\tilde{\Phi}\rangle \langle \tilde{\Phi}|\textbf{s}\rangle \Sigma(\textbf{s})}
\end{array}
\end{equation}
With Eq.~(\ref{PhiTil}) we thus have
\begin{equation}\label{Taur}
\tau_R=|\kappa(n)|^2 \sum_{\textbf{s}} I \otimes \left( \prod_{k=1}^n B_{s_k}\right)  |L\rangle \langle L |\, I \otimes \left(\prod_{l=n}^1 B_{s_l}^\dagger\right).
\end{equation}
We may rewrite this in a simpler form,
\begin{equation}\label{Channel}
\tau_R = |\kappa(n)|^2 I\otimes {\cal{L}}^{n}(|L\rangle\langle L|),
\end{equation}
where ${\cal{L}}^{n}$ denotes the $n$-fold iteration of ${\cal{L}}$, ${\cal{L}}^{n} = \underbrace{{\cal{L}} \circ {\cal{L}} \circ .. \circ {\cal{L}}}_{n\;\text{times}}$, and the channel ${\cal{L}}$ is
\begin{equation}\label{Ldef}
{\cal{L}}(\rho) = \sum_{i=0}^{d-1} B_i \rho B_i^\dagger.
\end{equation}
{By our assumption of uniqueness of the SPTO ground state, the tensors $\{A[i]\}$ satisfy the injectivity condition; See \cite{SPT1},\cite{SPT3}. Then, the matrices $\{B_i\}$ also satisfy injectivity, and the fixed point of the channel ${\cal{L}}$ of Eq.~(\ref{Ldef}) is therefore unique \cite{MPS}. }

By adjusting the factor $\kappa(n)$ in Eqs.~(\ref{Phi}), (\ref{PhiTil}), the normalization for the channel ${\cal{L}}$ may always be chosen such that the largest eigenvalue $\lambda_0$ of ${\cal{L}}$ is
\begin{equation}\label{lambda0}
\lambda_0 = 1.
\end{equation}
We adopt this convention in the following. The correlation length $\xi$ of the states $|\Phi\rangle$, $\tilde{\Phi}\rangle$ is then given by $\lambda_1$, the second-largest eigenvalue of ${\cal{L}}$, $\xi:=-1/\ln \lambda_1$. The unique fixed point of ${\cal{L}}$ is closely approximated after $n$-fold iteration of the channel ${\cal{L}}$ if $n/\xi \gg 1$,
\begin{equation}\label{wirefix}
{\cal{L}}^{n}(\rho) \approx \nu_\rho\, \rho_\text{fix},
\end{equation}
with $\nu_\rho\in \mathbb{R}_+$, for all states $\rho$ of the junk system. Therefore, if the virtual system is in a tensor product state $\sigma \otimes \rho$ between the logical and the junk subsystem prior to the action of a channel $I\otimes {\cal{L}}^{n}$, then the state afterwards is $\sim \sigma \otimes \rho_\text{fix}$. Thus, the tensor product structure of the state of the virtual system is preserved, and the junk part is driven towards a unique Hermitian fixed point state $\rho_\text{fix}$. This is an important computational ingredient which will be employed in the implementation of unitary gates and measurement.

In addition, even if the joint state between junk and the logical subsystems does not factorize initially, after the application of the oblivious wire it does. 
\begin{Lemma}\label{factor}
Consider a channel ${\cal{L}}$ of Eq.~(\ref{Ldef}) derived from an  {injective} resource state $|\Phi\rangle$ of Eq.~(\ref{Phi}), with left boundary state $|L\rangle$. For any such state it holds that
\begin{equation}
\lim_{n\rightarrow \infty} I\otimes {\cal{L}}^{n}(|L\rangle \langle L|) = \sigma \otimes \rho_\text{fix},
\end{equation}
for some Hermitian operator $\sigma$ depending on $|L\rangle$. Further, $\sigma \otimes \rho_\text{fix}$ is positive semidefinite.
\end{Lemma}\medskip

{\em{Proof of Lemma~\ref{factor}.}} The channel ${\cal{L}}$ has the following two properties. First, it is a linear map,
\begin{equation}\label{Lin}
{\cal{L}}(c\, A + d\, B) = c\, {\cal{L}}(A) + d\, {\cal{L}}(B),\;\; \forall c,d \in \mathbb{C}.
\end{equation}
{Second, under the condition of injectivity it holds that
\begin{equation}\label{Bfix}
\lim_{n\rightarrow \infty}{\cal{L}}^{n}(X) = \nu_X\, \rho_\text{fix},
\end{equation}
with $\nu_X\in \mathbb{C}$, for all operators $X$ \cite{MPS} (also see \cite{QETP}; Sec. 8.2).}

We may now write the state $|L\rangle$ of the left boundary in its Schmidt decomposition between the logical and the junk system,
$
|L\rangle = \sum_j\sqrt{\lambda_j} \, |\phi_j\rangle \otimes |\psi_j\rangle, 
$
where $\{\phi_j\}$, $\{\psi_j\}$ are ONBs of the logical and the junk system, respectively, and $\{\lambda_j\}$ are the non-zero eigenvalues of the reduced density matrix. Correspondingly,
$
|L\rangle \langle L| =  \sum_{i,j} \sqrt{\lambda_i\lambda_j^*}  |\phi_i\rangle \langle \phi_j|\otimes |\psi_i\rangle \langle \psi_j|
$.
We thus have
$$
\begin{array}{ll}
\multicolumn{2}{l}{\displaystyle{\lim_{n\rightarrow \infty}I\otimes {\cal{L}}^{n}(|L\rangle \langle L|)} =}\vspace{1mm} \\ 
&= \displaystyle{\sum_{ij} \sqrt{\lambda_i\lambda_j^*} |\phi_i\rangle \langle \phi_j| \lim_{n\rightarrow \infty}\otimes {\cal{L}}^{n}(|\psi_i\rangle \langle \psi_j|)}\vspace{1mm}\\
&= \displaystyle{\left[  \sum_{ij} \sqrt{\lambda_i\lambda_j^*} \,  \nu(n)_{|\psi_i\rangle\langle\psi_j|}  |\phi_i\rangle \langle \phi_j| \right]\otimes \rho_\text{fix}}\vspace{1mm}\\
&= \sigma\otimes \rho_\text{fix}
\end{array}
$$
Therein, we used linearity Eq.~(\ref{Lin}) in the first line, and Eq.~(\ref{Bfix}) in the second line. 

{Further, with Eq.~(\ref{Ldef}), ${\cal{L}}(X^\dagger) = ({\cal{L}}(X))^\dagger$, $\forall X$, and thus with Eq.~(\ref{Bfix}) it follows that $\nu_{X^\dagger}=\nu_X^*$. Hence $\sigma$ is Hermitian. Finally, since ${\cal{L}}$ is completely positive and $|L\rangle \langle L|$ is positive semidefinite, so is $\sigma \otimes \rho_\text{fix}$.} $\Box$
\medskip

\noindent
Thus, we obtain 
\begin{CP}
The virtual system can be prepared in a tensor product state $\tau=\sigma\otimes \rho_\text{fix}$, where the junk system is in a defined fixed point state $\rho_\text{fix}$.
\end{CP}

\subsection{Interpretation of ``adding computational paths''}\label{F/D}

In the oblivious wire construction---Procedure III in Section~\ref{FW}---we have added computational paths corresponding to different measurement records $\textbf{s}$ by ``forgetting'' those measurement records (after correction for the byproduct operators $\Sigma(\textbf{s})$). The purpose of this section is to clarify that ``forgetting'' of this classical information is a meaningful operation.

The simplest way to justify the notion of ``forgetting information'' is to consider a distributed scenario for computation, involving two parties. In this scenario, the primary party, Alice, does not need to forget classical information, but  never learns it in the first place.

Suppose, the above Procedure III is outsourced by Alice, the computing party, to some subcontractor Bob. The protocol between Alice and Bob is as follows. (i) Alice sends Bob the state $|\tilde{\Phi}\rangle$, (ii) Bob implements Procedure II, (iii) Bob sends back the right boundary system (which is a physical particle, since $|\tilde{\Phi}\rangle$ has been used instead of $|\Phi\rangle$). Bob does not say which measurement outcomes he obtained, and discards the measured spins (they are being traced over, from Alice's perspective).

How is Alice supposed to represent the quantum state of the right boundary system that she receives from Bob?---Since Alice has no information about the measurement outcomes obtained by Bob, from her perspective, the operation implemented by Bob is the channel with Kraus operators Eq.~(\ref{Alice}), leading to the state $\tau$ of the right boundary system given in Eq.~(\ref{Channel}).

However, in the intended computational scenario there is no second party. Rather, a single party is executing the computation in her lab. An interpretation for adding computational paths applicable to the one-party setting is given in Appendix~\ref{1P}.

\section{Unitary gates}\label{iG}

Next we show how to implement a single unitary about an infinitesimal rotation angle $d\alpha$ by measurement of the leftmost qudit in the chain, assuming the junk system is already in its fixed point state $\rho_\text{fix}$. For this purpose, we change slightly the computational setting from the previous section. The resource state is now mixed, and its mixedness comes from a mixed state in the left boundary condition. Namely, we consider the resource state of $n+1$ spins plus the right boundary system
\begin{equation}\label{TilPhiMixed}
\tilde{\Phi}_\text{fix}(\sigma) =  |\kappa(n)|^2 \sum_{\textbf{i},\textbf{j}}  \left( A[\textbf{i}] \left( \sigma \otimes \rho_\text{fix} \right) A[\textbf{j}]^\dagger  \right) \otimes |\textbf{i}\rangle\langle \textbf{j}|,
\end{equation}
where $A[\textbf{i}]:=A[i_{n+1}]\cdot .. \cdot A[i_1]$.

\subsection{Rotations about small angles}
 
To perform a rotation about an infinitesimal angle, we measure the first spin in a basis which slightly deviates from the basis that implements the wire, namely ${\cal{B}}(d\alpha) = \{|0'\rangle, |1'\rangle, |2\rangle,.., |m-1\rangle\}$, where
\begin{equation}\label{Bas1}
\begin{array}{rcl}
	|0'\rangle &=&  |0\rangle + d\alpha \,|1\rangle,\\
	|1'\rangle &=&  |1\rangle - d\alpha \, |0\rangle,
\end{array}	
\end{equation}
where $d\alpha \in \mathbb{R}$. The remaining $n$ spins are measured in the wire basis, and the measurement outcomes are forgotten as soon as the accumulated byproduct operator $\Sigma(\textbf{s})$ are reversed on the right boundary, {including the byproduct operator caused by the measurement of the first particle in the basis Eq.~(\ref{Bas1})}.

In complete analogy with Eq.~(\ref{Taur}), the effect of the measurements and byproduct operator reversion can be described on the level of the virtual system,
\begin{equation}
\sigma \otimes \rho_{\text{fix}} \longrightarrow  {\cal{T}}_{s_1}(\sigma \otimes \rho_{\text{fix}}),
\end{equation}
where ${\cal{T}}_{s_1}$ is the channel on the virtual system induced by the measurement of the first spin, with outcome $s_1$, followed by an oblivious wire ${\cal{L}}^{n}$. We now investigate the channels ${\cal{T}}_{s_1}$.

First, assume the outcome of the measurement is $0'$. The corresponding matrix $A[0']$ acting on the correlation space is 
$A[0'] = A[0] + d\alpha\, A[1] = C_0\left( I \otimes B_0 + d\alpha\, C_0^{-1} C_1 \otimes B_1 \right)$.
Henceforth, we denote 
$$
C_0^{-1}C_1=:C.
$$
Given a left boundary state that is factorized between logical system and junk system, $\sigma \otimes \rho_\text{fix}$, with the junk system in its fixed point state, the effect of the measurement of the first spin on the correlation system is, up to leading order in $d\alpha$,
$$
\begin{array}{rcl}
\sigma \otimes \rho_\text{fix} &\longrightarrow& \displaystyle{\sigma \otimes B_0 \rho_\text{fix} B_0^\dagger +}\\
&&\displaystyle{+ d\alpha \left(C\sigma \otimes B_1\rho_\text{fix} B_0^\dagger + \sigma C^\dagger \otimes B_0\rho_\text{fix}B_1^\dagger \right).}
\end{array}
$$
We now follow up with an oblivious channel ${\cal{L}}^{n}$, such that the junk system is back to its fixed point state. To describe the resulting state of the virtual system, it us useful to define the parameters $\nu_{ij}\in \mathbb{C}$ via
{
\begin{equation}\label{nu_def}
\lim_{n \rightarrow \infty}{\cal{L}}^{n} (B_i \rho_\text{fix} B_j^\dagger) = \nu_{ij}\, \rho_{\text{fix}},
\end{equation}
for all $i,j=0,..,d-1$. Since $\rho_\text{fix}$  is Hermitian and $({\cal{L}}(X))^\dagger = {\cal{L}}(X^\dagger)$ for all $X$, we have
\begin{equation}\label{Hermit}
\nu_{ij}^* =\nu_{ji}, \;\forall i,j.
\end{equation}
Furthermore, with Eq.~(\ref{lambda0}) it follows that
\begin{equation}\label{Norm}
\sum_{i=0}^{d-1} \nu_{ii} =1.
\end{equation}}
Now, the combined action of the measurement with outcome $0'$ and a subsequent oblivious channel  is $\sigma \otimes \rho_\text{fix} \longrightarrow {\cal{T}}_0(\sigma \otimes \rho_\text{fix})$, with
$$
{\cal{T}}_0(\sigma \otimes \rho_\text{fix}) = \nu_{00}\, \sigma \otimes \rho_\text{fix}
+ \,d \alpha \big(\nu_{10}\, C\sigma +\nu_{01} \,\sigma C^\dagger\big) \otimes \rho_\text{fix}.
$$
Thus, we are back to a disentangled logical subsystem, which was the goal.

Let us now examine which transformation was enacted on the logical system. For this purpose, we may split the term $\sim d\alpha$ in the above expression for the channel ${\cal{T}}_0$ into a commutator and an anti-commutator part,
$$
\begin{array}{rcl}
{\cal{T}}_0(\sigma \otimes \rho_\text{fix}) &=& \displaystyle{\nu_{00}\, \sigma \otimes \rho_\text{fix}+}\vspace{1mm}\\
&& \displaystyle{+ \frac{d \alpha}{2} \left[\nu_{10}\, C -\nu_{10}^* \,C^\dagger,\sigma\right] \otimes \rho_\text{fix} +}\vspace{1mm}\\
&&\displaystyle{+ \frac{d \alpha}{2} \left\{\nu_{10}\, C +\nu_{10}^* \,C^\dagger,\sigma\right\}\otimes \rho_\text{fix}.}
\end{array}
$$
The commutator term generates unitary evolution and is thus desirable, while the anti-commutator term generates non-unitary stretching which is undesirable. In general, both parts are present.\medskip

We now repeat the above calculation for the outcome $1'$. {After reversal of all byproduct operators on the right boundary system,} the action on the left boundary condition now is $|L\rangle \longrightarrow \left(I\otimes B_1 - d\alpha \,C^\dagger \otimes B_0\right) |L\rangle$. This holds because $C_1^{-1} C_0 = C^{-1}=C^\dagger$, which is a consequence of the fact that all byproduct operators in the wire basis are unitary, by assumption. We find that  $\sigma \otimes \rho_\text{fix} \longrightarrow {\cal{T}}_1(\sigma \otimes \rho_\text{fix})$, with
$$
{\cal{T}}_1(\sigma \otimes \rho_\text{fix}) = \nu_{11}\,\sigma \otimes \rho_\text{fix}
-\,d\alpha\big(\nu_{01}\,C^\dagger \sigma + \nu_{10}\, \sigma C\big) \otimes \rho_\text{fix}.
$$
Again, the linear order in $d\alpha$ has a unitary commutator part and a non-unitary anti-commutator part.  Expanding the above expression,
$$
\begin{array}{rcl}
{\cal{T}}_1(\sigma \otimes \rho_\text{fix}) &=& \displaystyle{\nu_{00}\, \sigma \otimes \rho_\text{fix}+}\vspace{1mm}\\
&& \displaystyle{+ \frac{d \alpha}{2} \left[\nu_{10}\, C -\nu_{10}^* \,C^\dagger,\sigma\right] \otimes \rho_\text{fix} +}\vspace{1mm}\\
&&\displaystyle{- \frac{d \alpha}{2} \left\{\nu_{10}\, C +\nu_{10}^* \,C^\dagger,\sigma\right\}\otimes \rho_\text{fix}.}
\end{array}
$$
Comparing the expressions for the action of the channels ${\cal{T}}_0$ and ${\cal{T}}_1$, we find that they have the same commutator part, with the same sign, and the same anti-commutator part, with opposite sign. Thus, upon adding the two channels ${\cal{T}}_0$ and ${\cal{T}}_1$, the anti-commutator part vanishes, and the evolution of the logical subsystem becomes purely unitary, up to linear order in $d\alpha$,
\begin{equation}\label{Ch2}
\begin{array}{rcl}
{\cal{T}}_0 + {\cal{T}}_1:\; \sigma \otimes \rho_\text{fix} &\longrightarrow& \displaystyle{(\nu_{00}+ \nu_{11}) \,\sigma \otimes \rho_\text{fix}
+}\vspace{1mm}\\
&&\displaystyle{d\alpha \left[\nu_{10} C-\nu_{10}^*C^\dagger ,\sigma\right] \otimes \rho_\text{fix}.}
\end{array}
\end{equation} 
Up to the norm factor $\nu_{00}+\nu_{11}$ this is now a unitary gate $U(d\alpha)$ with rotation angle $\sim d\alpha$,  
\begin{equation}\label{Rot1da}
U(d\alpha) = \exp\left(i\, d\alpha  \frac{\nu_{10} C-\nu_{10}^* C^\dagger}{i(\nu_{00}+\nu_{11})}\right),
\end{equation}
conditioned upon the measurement outcome 0 or 1 being obtained.

As discussed in Section~\ref{F/D}, ``adding together'' of the two channels ${\cal{T}}_0$ and ${\cal{T}}_1$ means that, if the measurement outcome $s_1$ was either 0 or 1, after the correction of the byproduct operator, the outcome is discarded, i.e. no longer available for any further processing. It is only remembered that one of 0 or 1 occurred.\medskip

The reason for conditioning the junk system is now clearly visible.  We have a reliable procedure for driving the junk system into a fixed point state $\rho_\text{fix}$. The state $\rho_\text{fix}$, and even the dimension of the Hilbert space it lives in,  are a priori unknown. But that doesn't matter. All that needs to be know about $\rho_\text{fix}$ are the $d(d-1)/2$ parameters $\nu_{ij}$, $j\geq i$. Those parameters can be measured.

The gate of Eq.~(\ref{Ch2}) is probabilistic but heralded. It simplifies our discussion to convert it into a deterministic gate, at the cost of reducing the rotation angle. This proceeds by adding in the channels ${\cal{T}}$, $i=2,\,..\, , d-1$. Individually, any such ${\cal{T}}_i$ acts as ${\cal{T}}_i: \sigma \longrightarrow \nu_{ii} \sigma$. And thus the net effect is that in Eq.~(\ref{Ch2}) $\nu_{00}+\nu_{11}$ in the rotation angle is replaced by $\sum_{i=0}^{d-1}\nu_{ii}=1$, i.e. the rotation angle is reduced. 

It may seem counter-intuitive that a probabilistic mixture of two unitaries should be a unitary. Yet, clearly there is no problem with this statement if the two unitaries are the same. Here, one of the two unitaries is the identity, and one is very close to the identity, deviating to linear order in $d\alpha$. The probabilistic mixture of the two is indeed not exactly unitary, but the difference shows up only to quadratic order in $d\alpha$, which we can discard in the present discussion.
\medskip

We now redo the above calculation for a continuous set of measurement bases $\tilde{{\cal{B}}}(d\alpha,\beta)$, with
\begin{equation}\label{BetaBas}
\begin{array}{rcl}
	|0(\beta)\rangle &=&  |0\rangle +  d\alpha\, e^{i\beta}\,|1\rangle,\\
	|1(\beta)\rangle &=&  |1\rangle  - d\alpha\, e^{-i\beta}\, |0\rangle,
\end{array}	
\end{equation}
and $d\alpha,\beta  \in \mathbb{R}$. We find that we can now realize the operations
\begin{equation}\label{Rot3}
U(d\alpha,\beta) = \exp\left(i\, d\alpha |\nu_{10}| \frac{\left(e^{-i(\beta+\delta)}C-e^{i(\beta+\delta)}C^\dagger\right)}{i}\right),
\end{equation}
where $\nu_{10}=|\nu_{10}|e^{-i\delta}$.

\subsection{Composition}\label{Compos}

We have so far shown how to implement a single unitary gate with small rotation angle. To accumulate finite rotation angles, MBQC-gate simulations need to be compsable. That is, we require that if the gates $T$ and $T'$ can individually be executed by the MBQC, then so can their compositions $TT'$ and $T'T$.

The key, as usual in MBQC, is that the randomness of measurement resulting in the byproduct operators can be counter-acted by the adjustment of measurement bases. If all byproduct operators $C_i\otimes I$ are elements of a projective representation $V(G)$ of the symmetry group $G$, acting on the logical part of the virtual system only, then byproduct operators can be propagated through the chain by virtue of the symmetry relation
\begin{equation}\label{PropRel}
\parbox{6cm}{\includegraphics[width=6cm]{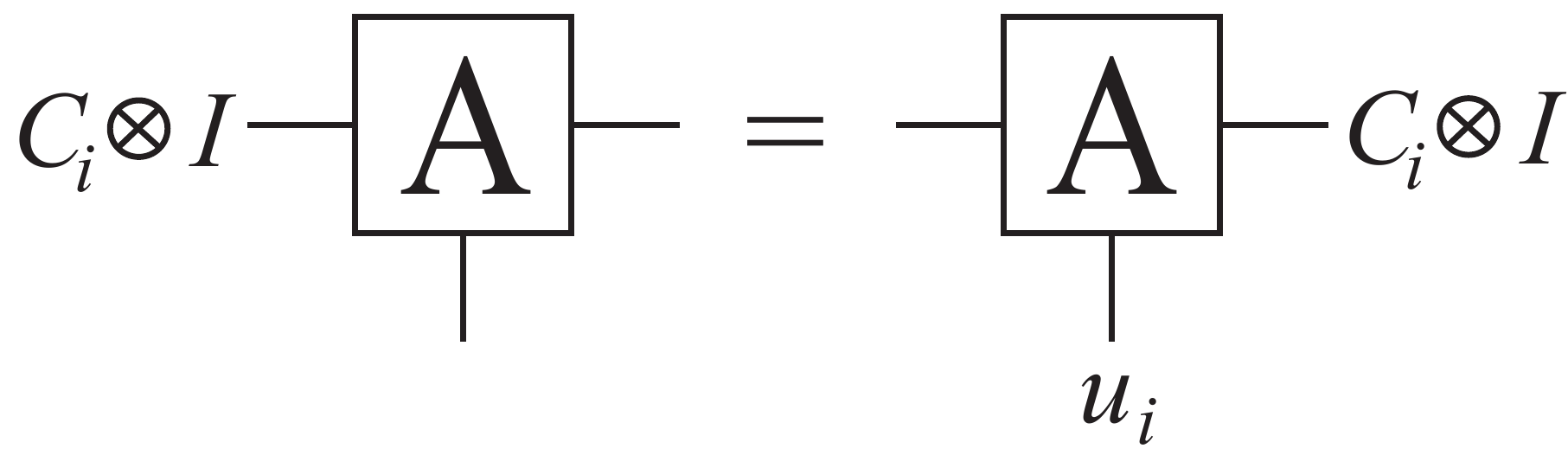}}.
\end{equation}
Therein, $u_i$ is in a unitary representation $U(G)$ of $G$, acting on the physical degrees of freedom. This situation ($C_i \in V(G),\,\forall i$) applies, for example, to maximally non-commuting factor systems of Abelian groups \cite{Else}.

Denote by $A(s,\beta)$ the MPS tensor representing the action on the virtual space caused by the measurement of a physical degree of freedom in the basis $\tilde{{\cal{B}}}(d\alpha, \beta)$ of Eq.~(\ref{BetaBas}), with outcome $s$. Further, denote by $A(\textbf{s},s,\beta):=(C(\textbf{s})\otimes I)A(s,\beta)(C(\textbf{s})^\dagger\otimes I)$ the MPS tensor obtained from $A(s,\beta)$ by conjugating under a byproduct operator $C(\textbf{s}) \otimes I$. With Eq.~(\ref{PropRel}), for all $\textbf{s}$ and all measurement angles $\beta$, there is a measurement basis ${\cal{B}}(\beta'(\textbf{s}))$ such that
\begin{equation}
A(\textbf{s},s,\beta) = A(s,\beta'(\textbf{s})).
\end{equation}
Hence, $A(\textbf{s},s,\beta)$ can be implemented.

Now recall that $\sum_s[A(s,\beta)]$ followed by oblivious wire ${\cal{L}}^{n}$ implements a channel $T(\beta)$ on states $\sigma_L \otimes \rho_\text{fix}$.  $\sum_s[A(\textbf{s},s,\beta)]$ followed by oblivious wire ${\cal{L}}^{n}$ thus implements a channel 
\begin{equation}\label{BasFlip}
T(\beta'(\textbf{s})) = C(\textbf{s}) T(\beta) C(\textbf{s})^\dagger
\end{equation}
on this set of states.

\begin{Lemma}\label{Concat}
Consider an injective MPS state $|\tilde{\Phi}\rangle$ of  Eq.~(\ref{PhiTil}) whose MPS tensors factorize w.r.t. a wire basis, $A[i]=B_i\otimes C_i$, and for all $i=0,..,d-1$ exists a $g\in G$ such that $C_i\otimes I=V(g)$. If two transformations $T(\alpha)$, $T'(\beta)$ can be implemented on $|\tilde{\Phi}\rangle$ then so can their composition $T'(\beta) T(\alpha)$.
\end{Lemma}
The proof of Lemma~\ref{Concat} is given in Appendix~\ref{conc}.

\subsection{Rotations about finite angles}

With Lemma~\ref{Concat}, we have so far established that we can execute all unitary gates of form Eq.~(\ref{Rot3}), but with finite rotation angles $\alpha$ replacing the infinitesimal angles $d\alpha$. 

Of course, there is nothing special in the choice of the basis elements 0 and 1 in the definition of $C:=C_0^{-1}C_1$ leading to Eq.~(\ref{Rot3}). We may replace the labels 0 and 1 by $i$ and $j$, $0\leq i, j\neq i \leq d-1$, giving rise to different operators $C=C_i^{-1}C_j$. This observation motivates Definition~\ref{Lie} in Section~\ref{SR}.

We have the following result.
\begin{Lemma}\label{SPTgates}
Consider a symmetry-protected phase of a group $G$ with the properties (i) the ground state is unique, (ii) there is a wire basis, and (iii) the byproduct operators $C_i$ are in a projective representation of $G$, $C_i\otimes I \in V(G)$.
Then, for all resource states $|\tilde{\Phi}\rangle$ of Eq.~(\ref{PhiTil}) derived from MPS ground states Eq.~(\ref{Phi}) in that phase, except a possible subset of measure zero, the unitary gates in $L({\cal{O}})$ can be arbitrarily closely approximated. 
\end{Lemma}

{\em{Proof of Lemma~\ref{SPTgates}.}} Whenever $\nu_{ij}\neq 0,\;\forall i,j=0,..,d-1$, all infinitesimal unitary gates in $L({\cal{O}})$ can be executed. Then, since the ground state is unique all across the SPTO  phase, MPS ground states are injective \cite{SPT1}, \cite{SPT3}, and Lemma~\ref{Concat} applies. Thus, with $\exp(i\,d\alpha \,A)\exp(i\,d\beta\,B)\approx \exp(i\,d\alpha\, A+i\,d\beta\,B)$, for all $A,B \in {\cal{O}}$, all rotations generated by linear combinations of observables in ${\cal{O}}$ can be realized. Furthermore,
$\exp(i\, d\alpha\, A)\exp(i\, d\alpha\, B)\exp(-i\, d\alpha\, A)\exp(-i\, d\alpha\, B)\approx \exp((d\alpha)^2[A,B])$, thus rotations  generated by any element in ${\cal{A}}({\cal{O}})$ can be realized, and hence all rotations in $L({\cal{O}})$. 

A value of $\nu_{ij}=0$, for some $i\neq j$, requires fine-tuning in the phase and thus only occurs for a set of states of measure zero. $\Box$\medskip

Thus, whenever wire is protected by symmetry, then a large group of gates and set of measurements can also be carried out all across the SPTO phase. These gates are not protected in the same way as wire is, and they have to be executed in a fashion reminiscent of the Zeno effect. Namely, the measurement bases  implementing unitary gates on the virtual system must always remain close to the wire basis, and finite rotation angles are accumulated over a large number of measurements. 

We restate Lemma~\ref{SPTgates}, without the conditions and qualifications, as the second computational primitive.
\begin{CP}
The logical part of the virtual system can be acted on by all unitary gates in $L({\cal{O}})$.
\end{CP}

\subsection{Operational cost of unitary evolution}

{In the above treatment of the unitary transformations, we have discarded terms of order $d\alpha^2$, and indeed, to quadratic order in $d\alpha$ deviations from unitarity arise. When finally composing gates, a rotation about a finite angle $\alpha$ is realized as a sequence of $N$ rotations about an angle $d\alpha = \alpha/N$. The error for each such rotation is thus of order $1/N^2$, and hence the error of the combined procedure is of order $1/N$. Thus, the error of the approximation can be improved by increasing the number $N$ of steps. An error $\epsilon$ requires order $1/\epsilon$ steps.}

\section{Measurement and initialization}\label{M}

{We have so far employed measurement only of the right boundary system in the support of $|\tilde{\Phi}\rangle$. This is unsatisfactory, since the presence of the right boundary system is only a transitional feature. It is removed in the final construction, which employs the original resource state $|\Phi\rangle$ of Eq.~(\ref{Phi}). Here, we describe a procedure of measurement of the virtual system. It is independent of the choice of the right boundary condition.} 

\subsection{Procedure for measurement}

For measurement of the virtual system, computational paths are no longer added, except in the oblivious wire. Furthermore, we choose the measurement bases a finite angle $\alpha$ away from the wire basis, ${\cal{B}}(\alpha)=\{|0'\rangle, |1'\rangle, |m'\rangle = |k\rangle,\; k\geq 2\}$, with
\begin{equation}\label{Balpha}
\begin{array}{rcl}
|0'\rangle &=& \cos \alpha |0\rangle+\sin \alpha |1\rangle,\\
|1'\rangle &=& -\sin \alpha |0\rangle +\cos \alpha |1\rangle.
\end{array}
\end{equation}
If the measurement outcome is $0'$ the corresponding action on the virtual system is thus $A'[0]=\cos\alpha\, A[0]+\sin\alpha\, A[1]= C_0(\cos\alpha\, I \otimes B_0+\sin \alpha \,C \otimes B_1)$. Thus, an initial state $\tau = \sigma\otimes \rho_\text{fix}$ transforms into
$$
\begin{array}{rcl}
\tau &\longrightarrow& \displaystyle{\cos^2\alpha\,  C_0 \sigma C_0^\dagger \otimes B_0 \rho_\text{fix} B_0^\dagger +}\vspace{1mm}\\
&& \displaystyle{+\sin^2\alpha\,  C_1 \sigma C_1^\dagger \otimes B_1 \rho_\text{fix} B_1^\dagger +}\vspace{1mm}\\
&&\displaystyle{+\sin \alpha\cos \alpha \, C_0 \sigma C^\dagger C_0^\dagger\otimes B_0 \rho_{\text{fix}}  B_1^\dagger +}\vspace{1mm}\\ 
&&\displaystyle{+\sin \alpha\cos \alpha \, C_0 C \sigma C_0^\dagger \otimes B_1 \rho_\text{fix} B_0^\dagger .}
\end{array}
$$
We now follow up the first measurement in the basis ${\cal{B}}(\alpha)$ with an oblivious wire and undoing of the byproduct operator on the right boundary system. The action of this chain of operations on the virtual system is ${\cal{T}}_0(\sigma \otimes \rho_\text{fix})=T_0(\sigma)\otimes \rho_\text{fix}$, with
\begin{equation}\label{T_0}
\begin{array}{rcl}
T_0(\sigma) &=& \cos^2\alpha \, \nu_{00} \,\sigma + \sin \alpha \cos \alpha\, \left(  \nu_{10}  C\sigma+ \nu_{10}^* \sigma C^\dagger \right)+\\ 
&&+ \sin^2\alpha \, \nu_{11} \, C\sigma C^\dagger  .
\end{array}
\end{equation}
First, for an eigenstate $|\phi_i\rangle$ of $C$, with $C|\phi_i\rangle = e^{i\phi_i}|\phi_i\rangle$, we find
\begin{equation}\label{T0gen}
\begin{array}{rcl}
T_0(|\phi_i\rangle \langle \phi_i |) &=& \cos^2\alpha\, \nu_{00} +\sin^2\alpha \,\nu_{11} +\\
&&+  |\nu_{10}|\sin 2\alpha\, \cos(\delta+\phi_i) |\phi_i\rangle \langle \phi_i|,
\end{array}
\end{equation}
where we have set $\nu_{10} =e^{i\delta}|\nu_{10}|$. 

\begin{figure}
\begin{center}
\includegraphics[width=6cm]{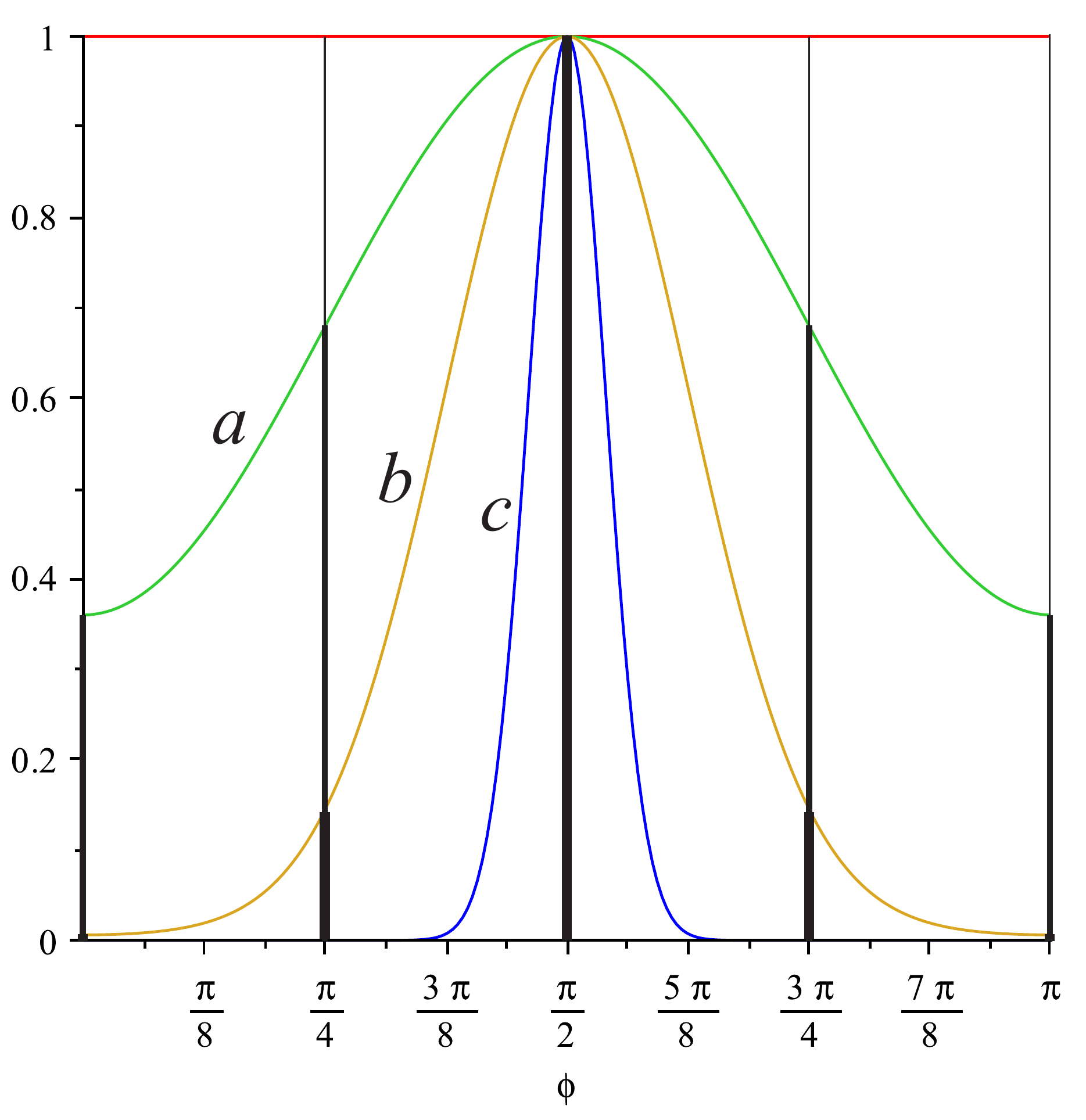}
\caption{\label{FiFu} Accumulated filter functions $f_0^{N_0}f_1^{N_1}$ (normalized). (a) $N_0=N_1=1$, (b) $N_0=N_1=5$, (c) $N_0=N_1=50$. The parameters for this plot are $\nu_{00}=\nu_{11}=1$, $\nu_{10}=0.8$, {$\alpha=\pi/4$}.}
\end{center}
\end{figure}

{We now consider the state $\sigma$ of the logical part of the virtual system in the eigenbasis ${\cal{B}}_C$ of $C$, 
$
{\cal{B}}_C=\{|\phi_i\rangle|\, C|\phi_i\rangle = e^{i\phi_i}|\phi_i\rangle \}
$, 
$$\sigma:=\sum_{i,j}\sigma_{ij}\,|\phi_i\rangle\langle \phi_j|.
$$
By the same argument as in Eq.~(\ref{T_0}) we find that, conditioned upon obtaining the outcome $k$ in the measurement basis ${\cal{B}}(\alpha)$ of Eq.~(\ref{Balpha}), the expansion coefficients $\sigma_{ij}$ of $\sigma$ are updated as
\begin{equation}\label{f-act}
\sigma_{ij} \longrightarrow \sigma'_{ij} = f_k(\phi_i,\phi_j)\sigma_{i,j}.
\end{equation}
We call the multipliers $f_k(\phi_i,\phi_j)$ filter functions, since they amplify or diminish some eigenstates over others. It suffices to consider the filter functions $f_k(\phi_i,\phi_i)$ multiplying the diagonal elements of $\sigma$. Up to a state-independent normalizations, those filter functions are
\begin{equation}\label{f}
\begin{array}{rcl}
f_0 (\phi_i,\phi_i) &=& \nu_{00} \cos^2\alpha + \nu_{11}\sin^2\alpha +\\
&& + |\nu_{10}| \sin (2\alpha)  \cos(\delta+\phi_i),\vspace{1mm}\\ 
f_1 (\phi_i,\phi_i) &=& \nu_{00} \sin^2\alpha +  \nu_{11}\cos^2\alpha -  \\
&& -|\nu_{10}| \sin(2\alpha)  \cos(\delta+\phi_i),\vspace{1mm}\\
f_k(\phi_i,\phi_i) &=& \nu_{kk},\; k\geq 2.
\end{array}
\end{equation} 
Each measurement of a physical spin in the basis ${\cal{B}}(\alpha)$ thus amounts to a  weak measurement of the logical part of the virtual system, in which some of the eigenstates of $C$ are relatively boosted while others are diminished. In general no state is completely extinguished in a single measurement.  
As the filter functions accumulate, they dim out all values of $\cos\phi$ except an increasingly narrow band. In the limit of a large number of repetitions, a projective measurement is achieved. See Fig.~\ref{FiFu} for illustration. }

If in a sequence of measurements the outcome $0'$ is obtained $N_0$ times and the outcome $1'$ is obtained $N_1$ times, the accumulated filter function $F_{N_0,N_1}(\phi)=f_0(\phi,\phi)^{N_0}f_1(\phi,\phi)^{N_1}$ as a function of $\phi$ reaches its maximum at an angle $\phi_\text{max}$ given by
\begin{equation}\label{frat}
\frac{f_0(\phi_\text{max},\phi_\text{max})}{f_1(\phi_\text{max},\phi_\text{max})} = \frac{N_0}{N_1}.
\end{equation}
In the limit of large $N_0$, $N_1$, the eigenvalue $e^{i\phi}$ of $C$ with $\phi$ closest to $\phi_\text{max}$ is the outcome of the measurement.

{With Eq.~(\ref{f}), we obtain an estimate for $\phi$ from the outcome frequencies $N_0$, $N_1$,
\begin{equation}\label{Mcos}
\begin{array}{rcl}
\cos(\delta+\phi) &=& \displaystyle{\frac{\sin^2\alpha}{\sin2\alpha}\frac{(N_0\nu_{00}-N_1\nu_{11})}{(N_0+N_1)|\nu_{10}|}+}\vspace{1mm}\\
&&   \displaystyle{ +\frac{\cos^2\alpha}{\sin2\alpha} \frac{(N_0\nu_{11} -N_1\nu_{00})}{(N_0+N_1)|\nu_{10}|}.}
\end{array}
\end{equation}  
The outcomes $k\geq 2$ of measurement in the basis ${\cal{B}}(\alpha)$ have no effect on the measurement of the virtual system.}\medskip

\begin{figure}
\begin{center}
\begin{tabular}{r}
\includegraphics[width=7cm]{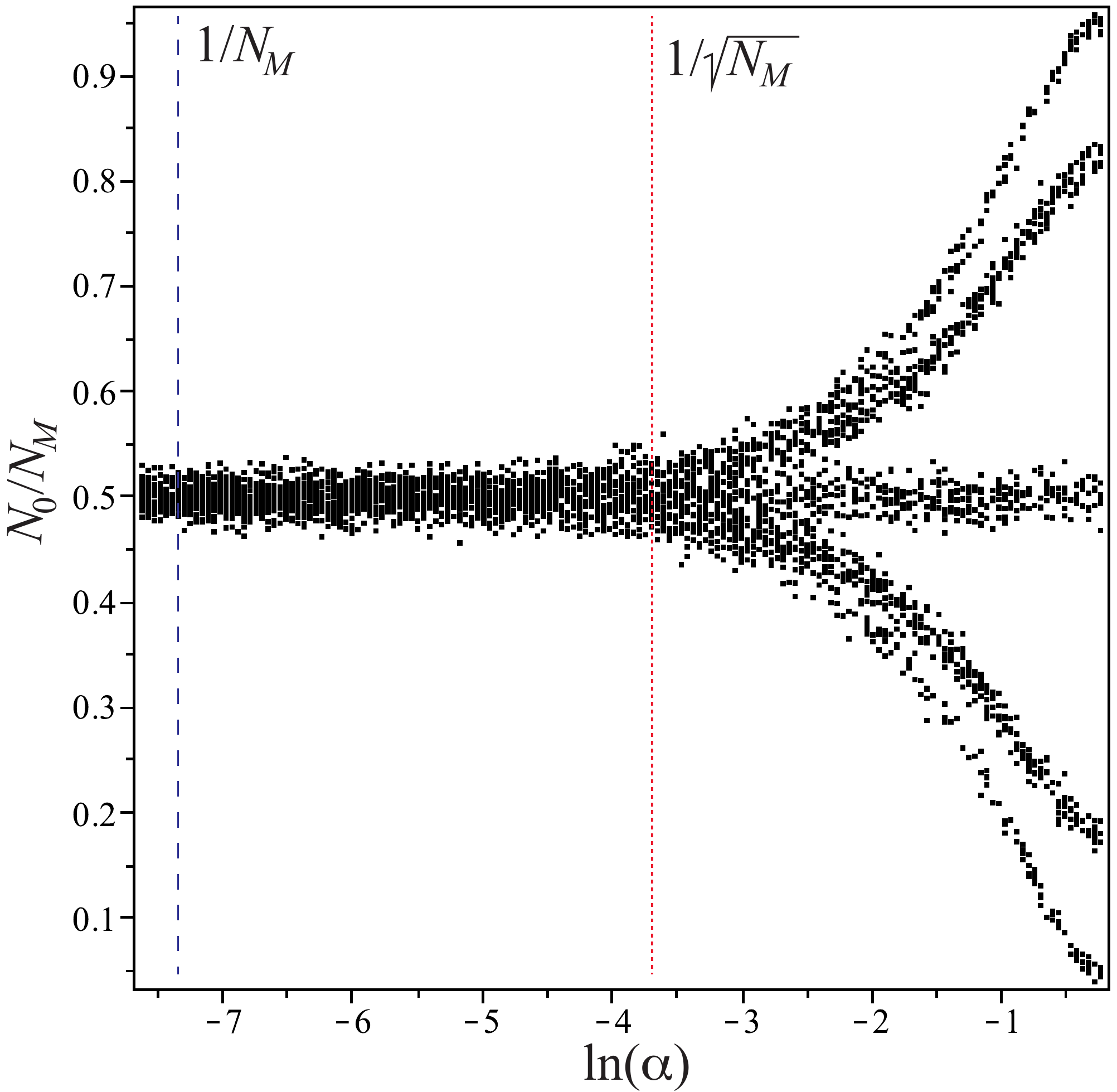} \\ \\
\includegraphics[width=7cm]{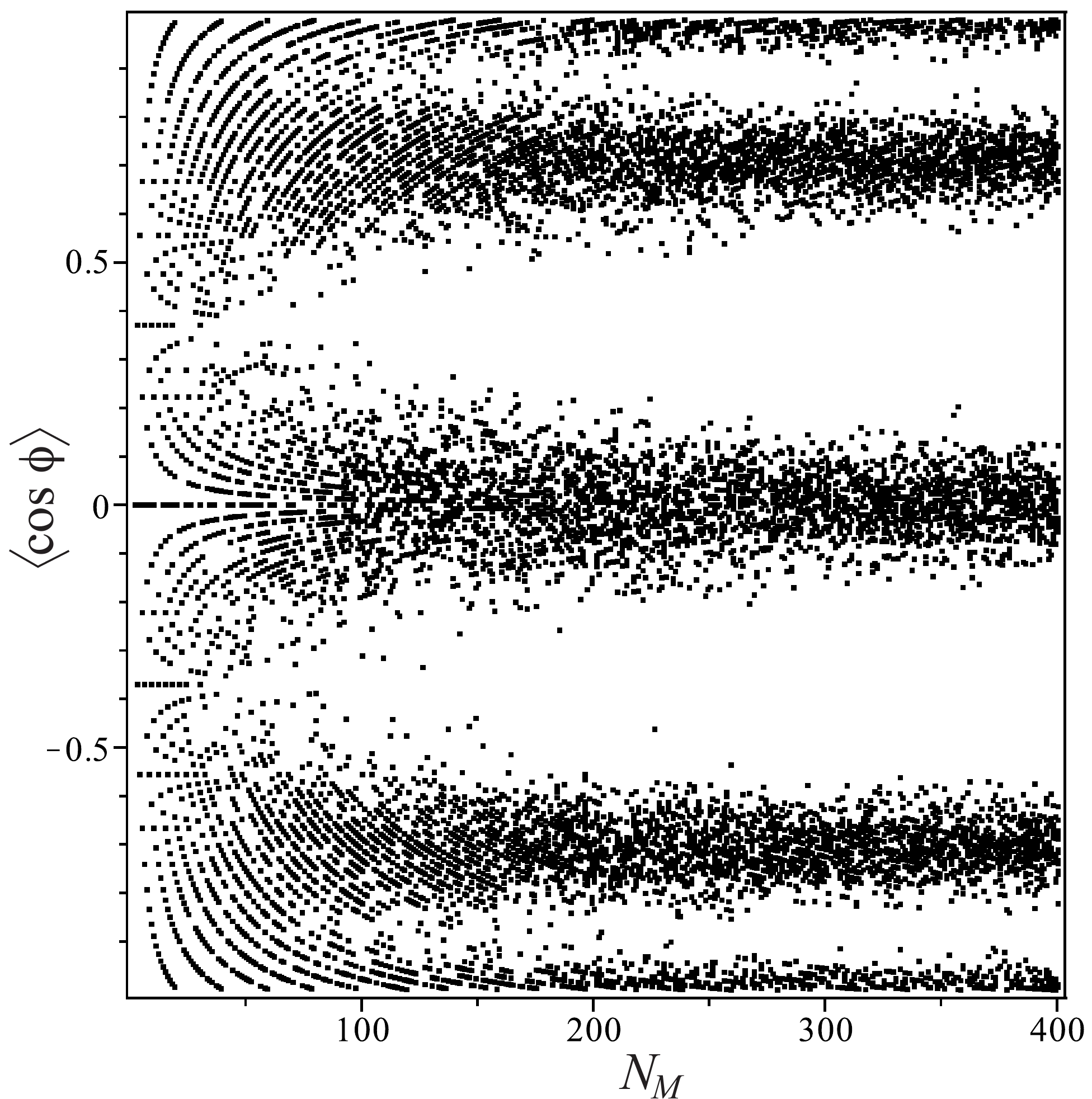} 
\end{tabular}
\caption{\label{means} Measurement of the logical subsystem of the virtual system. (top) Changeover between unitary evolution and measurement. For unitary evolution of the logical subsystem, the angle $\alpha$ in ${\cal{B}}(\alpha)$ is chosen $\sim 1/N_M$. Unitary evolution turns into measurement at $\alpha \approx1/\sqrt{N_M}$. The measurement is optimal for $\alpha =\pi/4$. In this plot, $N_M=1600$. (bottom) Onset of the large-$N$ limit in measurement. Each dot in the plot is the estimate of $\cos \phi_k$, where $e^{i\phi_k}$ are the eigenvalues of the measured observables $C$. Each such estimate is based on a sequence of $N_M$ individual weak measurements. In this example,  $\phi_k=\pi k/4$, $k\in \mathbb{Z}_8$. The parameters for both plots are $\nu_{00}=\nu_{11}=1$, $\nu_{10}=0.9$, and the state of the logical subsystem before measurement is completely mixed.}
\end{center}
\end{figure}

Unitary evolution and  measurement of the logical part of the virtual system are illustrated in Fig.~\ref{means}. In panel (a), the changeover between unitary evolution and measurement is shown. To implement a unitary gate with rotation angle ${\cal{O}}(1)$ on the logical part of the virtual system, a large number of consecutive spins in the chain is measured such that the outcome 0 is found $N_0$ times and 1 is found $N_1$ times, and $N_M=N_0+N_1$. The angle $\alpha$ specifying the measurement basis ${\cal{B}}(\alpha)$ is chosen $\sim 1/N_M$. The measurement basis is  thus very close to the wire basis.  Unitarity on the virtual system is beginning to be violated around  $\alpha = 1/\sqrt{N_M}$. The measurement is optimal for $\alpha =\pi/4$. Panel (b) shows the onset of the large-$N$ limit in measurement. Each dot in the plot is the estimate of $\cos \phi_k$, where $e^{i\phi_k}$ are the eigenvalues of the measured observables $C$. (In this example,  $\phi_k=\pi k/4$, $k\in \mathbb{Z}_8$). The measurement procedure consists of $N_M$ weak measurements in a row, each at an angle $\alpha=0.5$ away from the wire basis. For small $N_M$ there are fewer points, because most estimates are outside the meaningful region of $|\cos \phi|\leq 1$. \medskip

The above weak measurement, estimating $\cos \phi +\delta$, has a degeneracy, since it cannot distinguish between the angles $\phi$ and $\phi'=-\phi-2\delta$. This degeneracy can be removed by performing a second sequence of measurements in the basis ${\cal{B}}^*(\alpha)=\{|0'\rangle,|1'\rangle, |2\rangle, ...\}$, with
\begin{equation}\label{BalphaPr}
\begin{array}{rcl}
|0'\rangle &=& \cos \alpha |0\rangle+i\sin\alpha |1\rangle,\\
|1'\rangle &=& i\sin \alpha |0\rangle +\cos \alpha |1\rangle.
\end{array}
\end{equation}
In this way, $\sin(\delta+\phi)$ is measured on the virtual system, and in combination with the above one obtains $e^{i(\phi+\delta)}$, hence $\phi$. We thus arrive at the following Lemma.

\begin{Lemma}\label{SPTmeas}
Consider a symmetry-protected phase of a group $G$ with the properties (i) the ground state is unique, (ii) there is a wire basis, and (iii) the byproduct operators $C_i$ are in a projective representation of $G$, $C_i\otimes I \in V(G)$.
Then, for all resource states $|\tilde{\Phi}\rangle$ of Eq.~(\ref{PhiTil}) derived from MPS ground states Eq.~(\ref{Phi}) in that phase, except a possible subset of measure zero,
projective measurement of the observables in the set  ${\cal{O}}$ can be arbitrarily closely approximated. 
\end{Lemma}
We restate this result, without the conditions and qualifications, as our third computational primitive.
\begin{CP}
All observables $A\in {\cal{O}}$ can be measured on the logical part of the virtual system.
\end{CP}
{For matters of efficiency of the measurement procedure, it is noted that in addition to measurement in the bases Eq.~(\ref{Balpha}) and (\ref{BalphaPr}) for the physical spins, leading to an estimate of $\cos(\phi+ \delta)$ and $\sin(\phi+ \delta)$ respectively, one may as well measure in the basis ${\cal{B}}(\alpha,\beta)$ defined by
\begin{equation}\label{Bab}
\begin{array}{rcl}
|0'\rangle &=& \cos\alpha |0\rangle + e^{i\beta} \sin\alpha |1\rangle,\\ 
|1'\rangle &=& \sin\alpha |0\rangle - e^{i\beta} \cos\alpha  |1\rangle.
\end{array} 
\end{equation}
Thereby, an estimate of $\cos(\phi+\delta+\beta)$ is obtained with the same efficiency. One may vary the angle $\beta$ to perform state tomography, or adjust $\beta$ to optimize the resolution of the measurement for estimating $\phi$ (choose $\beta$ such that $\cos(\phi+\delta+\beta)\approx 0$).}\medskip

\subsection{Born rule for the logical subsystem}

In the above, we have demonstrated that for a measurement of the logical subsystem of the virtual system, the post-measurement state is related to the measurement record in the way described by quantum mechanics, i.e., it is an eigenstate of the measured observable. To complete our discussion of measurement of the logical subsystem, we need relate the measurement record to the state before measurement.
\begin{Lemma}\label{BR}
In the measurement of an observable $C$ on the state $\sigma$ of the logical subsystem of the virtual system, the probability $p_C(i)$ of obtaining the post-measurement state $|\phi_i\rangle$ is given by the Born rule,
\begin{equation}
p_C(i) = \langle \phi_i |\sigma |\phi_i\rangle. 
\end{equation}
\end{Lemma}\medskip

{\em{Proof of Lemma~\ref{BR}.}} We consider the channel resulting from combining all computational paths of an individual weak measurement, as described in Eq.~(\ref{f-act}). With Eqs.~(\ref{Norm}), (\ref{f}) we find that
$$
\sigma_{ii} \longrightarrow \sum_{k=0}^{d-1} f_k(\phi_i,\phi_i) = \sigma_{ii},\;\forall i=1,..,D.
$$
Hence, all diagonal elements of $\sigma$ in the eigenbasis of $C$ are the same before and after the channel. This also holds for any number of iterations of the channel, 
\begin{equation}\label{befaf}
\sigma_{ii,\prec} = \sigma_{ii,\succ}.
\end{equation}
Now consider a sequence of measurements, with record $\textbf{k}=(k_1,..,k_m)$, that is sufficiently long to project  the logical state. Denote by $\textbf{k} \rightarrow i$ that the measurement record $\textbf{k}$ is interpreted as outcome $i$, and by $p_C(\textbf{k})$ the probability for the outcome $\textbf{k}$. The matrix element $\sigma_{ii}(\textbf{k})$ of the logical state conditioned on the outcome $\textbf{k}$ is
$$
\sigma_{ii}(\textbf{k}) = \left\{ \begin{array}{ll}  1, & \text{if } \textbf{k} \rightarrow i,\\ 0, & \text{if } \textbf{k} \not \rightarrow i. \end{array}\right.
$$
Then, for the matrix element $\sigma_{ii}(\succ)$ after measurement it holds that
$$
\begin{array}{rcl}
\sigma_{ii,\succ} &=& \sum_\textbf{k} \sigma_\textbf{ii}(\textbf{k}) p_C(\textbf{k})\\
&=& \sum_{\textbf{k}|\textbf{k}\rightarrow i}p_C(\textbf{k})\\
&=& p_C(i).
\end{array}
$$
For the matrix element $\sigma_{ii,\prec}$ before measurement it holds by definition that $\sigma_{ii,\prec} = \langle \phi_i|\sigma| \phi_i\rangle$. Hence, with Eq.~(\ref{befaf}), $\langle \phi_i|\sigma| \phi_i\rangle=p_C(i)$. $\Box$

\subsection{Operational cost of measurement}

Measurement is most effective for the choice $\alpha=\pi/4$ in the measurement basis  Eq.~(\ref{Balpha}). Then, Eq.~(\ref{Mcos}) for the estimate of $\cos (\phi+\delta)$ simplifies to
$$
\cos(\phi+\delta) = \frac{N_0-N_1}{N_0+N_1}\frac{\nu_{00}+\nu_{11}}{2|\nu_{01}|}.
$$
We first discuss the case where, after a couple of measurement rounds, it is found that $\cos(\phi+\delta) \approx 0$. In this region, the conversion of $\cos(\phi+\delta)$ into $\phi$ is the most reliable. Assuming that the uncertainty in the constants $\nu_{00}, \nu_{11}, |\nu_{01}|$ and $\delta$ is negligible, the experimental uncertainty $\Delta \phi$ of $\phi$ is $\Delta \phi = 1/\sqrt{N_M}\cdot (\nu_{00}+\nu_{11})/4|\nu_{01}|$.

Denote by $\Delta$ the smallest distance between two consecutive eigenvalues of the measured observable $C_0^{-1}C_1$ on the unit circle. To reach an accuracy $\Delta \phi = \epsilon \Delta$, with $\epsilon$ a small parameter, the number $N$ of measurement steps used for the measurement of the logical subsystem is $N_M /(\nu_{00}+\nu_{11})$, and thus\begin{equation}\label{Nm}
N = \frac{1}{(4\epsilon \Delta)^2} \frac{(\nu_{00}+\nu_{11})}{|\nu_{01}|^2}.
\end{equation}
In the remaining case, $\cos(\phi + \delta)\not\approx 0$, the procedure is as follows. First, $\phi+\delta$ is measured with low accuracy, in only a few rounds of measurement. Then, the further measurements proceed in a basis ${\cal{B}}(\pi/4,\beta)$ of Eq.~(\ref{Bab}) where $\beta$ is chosen such that  $\cos(\phi+\delta+\beta)\approx 0$. This results in a precision measurement of  $\cos(\phi+\delta-\beta)$, with the same operational overhead as in Eq.~(\ref{Nm}).

\subsection{Initialization}

Measurement can also be used for initialization of the logical system. To initialize in a fixed reference state, one may measure a suitable byproduct operator $C$, and subsequently apply a conditional unitary.

\section{Reverting to the boundary conditions of Eq.~(\ref{Phi})}\label{Cc}

In this section we convert to the resource state with the original boundary conditions, and thereby complete the construction. We show that the state $|\tilde{\Phi}\rangle$ of Eq.~(\ref{PhiTil})  with the modified boundary condition can be replaced by the original resource state $|\Phi\rangle$ of Eq.~(\ref{Phi}). The basic reason for why this works is that the computation is completed with a measurement somewhere in the bulk of the chain, sufficiently far from the boundary. Then, because the correlation length is typically finite, the choice of boundary condition does not matter. Below we formalize this intuition.

\begin{Lemma}\label{Bchoice}
If an MBQC can be performed with a quantum state $|\tilde{\Phi}\rangle$ of Eq.~(\ref{PhiTil}) then it can also be performed by the quantum state $|\Phi\rangle$ of Eq.~(\ref{Phi}), which differs from $|\tilde{\Phi}\rangle$ in the choice of the boundary condition at the right end of the chain.
\end{Lemma}

{\em{Proof of Lemma~\ref{Bchoice}.}} The proof splits into two parts. First we show that standard open boundary conditions are fine if the state associated with the right boundary is $I/D \otimes \overline{\rho}_\text{fix}$, where $\overline{\rho}_\text{fix}$ is the fixed point of the channel $\overline{\cal{L}}:=\sum_s [B^\dagger_s]$ with the largest eigenvalue $\overline{\nu}$, and $D$ is the dimension of the logical system. Second, we show how to prepare that state on the right boundary, starting from any boundary state $|R\rangle$.

(i) {The main point is that if the logical state of the right boundary system is $\sigma\sim I$, then the forward-propagated byproduct operator $\Sigma(\textbf{s})$ annihilates on the right boundary, $\Sigma(\textbf{s})\, I \, \Sigma(\textbf{s})^\dagger=I$. This is the counterpart to the active correction of the byproduct operator $\Sigma(\textbf{s})$ in case of the boundary conditions Eq.~(\ref{PhiTil}). In case of the boundary conditions Eq.~(\ref{Phi}) no active intervention is necessary.} 

For notational simplicity, we consider a sequence $T_{C,s}' \circ T$ of only two operations, where  $T$ is a unitary and $T'_{C,s}$ is a measurement of the observable $C$ with outcome $s$. W.l.o.g. we assume that the input state of the virtual system prior to these gates is $\sigma \otimes \rho_\text{fix}$. Any such input state can be prepared, as we have previously shown. We now consider the probabilities $p(s)$ and $p'(s)$ for obtaining the outcome $s$ if the computation is run on an input state $|\tilde{\Phi}\rangle$ and $|\Phi\rangle$, respectively. The probabilities are
$$
\begin{array}{rcl}
p(s) & = &\text{Tr} \left(T'_{C,s} \circ T (\sigma)\right)  \,\text{Tr}  (\rho_\text{fix}) \times c,\\
p'(s) &= &\text{Tr}  \left(T'_{C,s} \circ T (\sigma)\right)  \,\text{Tr}  (\rho_\text{fix} \overline{\rho}_\text{fix}) \times c'/D,
\end{array}
$$
where $c, c'$ are normalization constants independent of $s$, and thus $p(s) \sim p'(s)$. Since $\sum_s p(s) = \sum_s p'(s) =1$,
$$
p(s) = p'(s),\;\;\forall s=0,\,..\,,d-1.
$$
Thus there is no effect of the right boundary on the measurement statistics, if the right boundary state $I/D \otimes \overline{\rho}_\text{fix}$ can be prepared.

(ii) The right boundary state $I/D\otimes \overline{\rho}_\text{fix}$ is prepared by running ``completely oblivious'' wire from the right boundary inwards. Completely oblivious wire is the same as oblivious wire, except that measurement outcomes are discarded without performing any correction anywhere. First, the state $I/D \otimes \overline{\rho}_\text{fix}$ is indeed a fixed point of the evolution $\overline{\cal{F}}=\sum_s[A(s)]$,
$$
\begin{array}{rcl}
\displaystyle{\sum_{s=0}^{d-1} A(s)^\dagger \left( \frac{I}{D} \otimes  \overline{\rho}_\text{fix}\right) A(s)} &=& \sum_{s=0}^{d-1} C_s^\dagger \frac{I}{D} C_s \otimes B_s^\dagger \overline{\rho}_\text{fix} B_s\\
 &=& I/D \otimes  \sum_{s=0}^{d-1} B_s^\dagger \overline{\rho}_\text{fix} B_s\vspace{1mm}\\
 &=& \overline{\nu} \left( I/D \otimes \overline{\rho}_\text{fix}\right).
\end{array}
$$
Second, the state $ I/D \otimes \overline{\rho}_\text{fix}$ is the eigenstate of $\overline{\cal{F}}$ with the largest eigenvalue. Suppose there was an eigenstate $\tau$ of $\overline{{\cal{F}}}$ with larger eigenvalue $\overline{\nu}_\tau$, $\overline{\nu}_\tau >\overline{\nu}$. Define $\rho(\tau):=\text{Tr}_L\tau$, where $\text{Tr}_L$ is the trace over the logical subsystem. By assumption we have
$\overline{\nu}_\tau\, \tau =\overline{\cal{F}}(\tau) =\sum_s C_s^\dagger \otimes B_s^\dagger \,\tau\, C_s \otimes B_s$. Now taking $\text{Tr}_L$ on both sides, and using the cyclicity of trace, we find
$$
\overline{\nu}_\tau\, \rho(\tau) = \sum_s B_s^\dagger \rho(\tau) B_s = \overline{\cal{L}}(\rho(\tau)).
$$
$\overline{\cal{L}}$ has a larger eigenvalue than $\overline{\nu}$. Contradiction. Thus, the largest eigenvalue of the channel $\overline{\cal{F}}$ is $\overline{\nu}$, the same as of the channel $\overline{\cal{L}}$ on the junk system alone. 

The largest eigenvalue of $\overline{\cal{F}}$ is typically non-degenerate, and $I/D \otimes \overline{\rho}_\text{fix}$ is thus the unique stable fixed point. It is closely approximated for a sufficient number of iterations of $\overline{\cal{F}}$, starting from any boundary condition $|R\rangle$. $\Box$\smallskip

{
{\em{Remark:}} Completely oblivious wire is equivalent to the respective spins being traced out. To implement it, all that needs to be done is to keep a sufficiently long runway of spins between the last spin used for measurement and the right boundary of the chain.}\medskip

{\em{Proof of Theorem~\ref{MT}.}} Lemma~\ref{SPTgates} provides the unitary gates in $L({\cal{O}})$ and  Lemma~\ref{SPTmeas} projective measurements of observables in the set ${\cal{O}}$, given the state $|\tilde{\Phi}\rangle$ of Eq.~(\ref{PhiTil}). With Lemma~\ref{Bchoice}, the resource state $|\tilde{\Phi}\rangle$  may be replaced by the resource state $|\Phi\rangle$ of Eq.~(\ref{Phi}) with standard open boundary conditions on both sides. $\Box$

\section{Example: $\mathbb{Z}_D\times \mathbb{Z}_D$ symmetry}

{As an example, we consider symmetry-protected topological order for the symmetry group  $G=\mathbb{Z}_D\times \mathbb{Z}_D$, and are interested in the phase corresponding to the maximally non-commuting factor system. The latter describes a unique projective representation $V(G)$ acting on the virtual system as $V(a\times b) = \tilde{V}(a\times b)\otimes I_\text{junk}$, where $a\times b\in \mathbb{Z}_D\times \mathbb{Z}_D$, $\tilde{V}(a\times b) = X^aZ^b$  and $X$, $Z$ are Heisenberg-Weyl operators in dimension $D$, defined by $X|z\rangle =|z+1 \mod D\rangle$, $Z|z\rangle =e^{2\pi i\, z/D}|z\rangle$. With Theorem~1 of \cite{Else} and the proof thereof, the MPS tensor $A$ can be written as 
$$
A[i] = \tilde{V}(g_i) \otimes B_{\text{junk},i} \
$$
where $g_i \in \mathbb{Z}_D\times \mathbb{Z}_D$.} 

The dimension $d$ of the physical spins may now be chosen $d=D^2$, which is compatible with a qudit cluster state. Then, the operators $C_i^{-1}C_j$, for $i,j\neq i \in 0,\,..\, ,D^2-1$ are, up to phases, also the Heisenberg-Weyl operators  in dimension $D$ minus the identity. There are thus $D^2-1$ linearly independent such operators. Therefore the set of Eq.~(\ref{Odef}), for $d=D^2$, also has $D^2-1$ linearly independent elements. Furthermore, all elements of this set are Hermitian and traceless. Hence they span the space of traceless Hermitian $D\times D$ matrices. With Lemma~\ref{SPTgates}, the set of gates realizable in this SPTO phase is thus $SU(D)$. Furthermore, by Lemma~\ref{SPTmeas}, all Heiseberg-Weyl operators can be measured in this phase.

\section{The matrix $[\nu_{ij}]$}

\subsection{Measurement}

Running of MBQC at any given point in the SPTO phase requires knowledge of the $d^2$ parameters $\nu_{ij}$ characterizing the fixed point state $\rho_\text{fix}$ of the junk system. These parameters can be estimated as follows. First,  a state $\sigma\otimes \rho_\text{fix}$ of the virtual system is prepared through oblivious wire (See computational primitive 1). This procedure does not require knowledge of the parameters $\nu_{ij}$. Then follows a sequence of measurements in the wire basis alternating with oblivious wire. The frequencies $N_k$ of obtaining the outcomes $k$ in measurement in the wire basis are, with Eqs.~(\ref{frat}) and (\ref{f}), related via
$$
\frac{\nu_{kk}}{\nu_{ll}}=\frac{N_k}{N_l}.
$$
In this way, also invoking the normalization condition Eq.~(\ref{Norm}), all diagonal elements $\nu_{kk}$ are measured.

Finally, the off-diagonal elements can be determined by implementing rotations Eq.~(\ref{Rot3}) and measuring the rotation angle, e.g. by change in the population of the various levels of the logical subsystem.

\subsection{Interpretation}

The operator $\nu:=\sum_{i,j=0}^{d-1} |i\rangle \nu_{ij} \langle j|$ can be given the interpretation of a (mixed) quantum state. With Eq.~(\ref{Hermit}), $\nu$ is Hermitian, and with Eq.~(\ref{Norm}), $\text{Tr}\,\nu=1$. Finally, $\nu$ is positive semidefinite.

The latter can be seen as follows. Consider a tensor product state $\sum_{ij} B_i \rho_\text{fix} B_j^\dagger \otimes |i\rangle \langle j|$ of the junk part of the virtual system and a physical degree of freedom, where $\{|i\rangle\}$ is an ONB of $\mathbb{C}^d$ (the Hilbert space of a physical degree of freedom). For any state $|\psi\rangle$ of the physical system, consider furthermore the expression
$$
\begin{array}{rcl}
\nu_\psi(n) &: =& \langle \psi | \text{Tr}_\text{junk}\left(I \otimes {\cal{L}}^n\left( \sum_{ij} |i\rangle \langle j| \otimes  B_i \rho_\text{fix} B_j^\dagger  \right)\right)|\psi\rangle\\
&=& \text{Tr}_\text{junk}\left( I \otimes {\cal{L}}^n \left(B_\psi \rho_\text{fix} B_\psi^\dagger\right)\right),
\end{array}
$$
with $B_\psi = \sum_i \langle \psi |i\rangle B_i$. With Lemma~\ref{factor}, $\rho_\text{fix}$ is either positive or negative semidefinite, and wlog. may be chosen positive semidefinite (hence, $\text{Tr}(\rho_\text{fix})>0$). Since $[B_\psi]$ (as a super-operator) is completely positive for all $|\psi\rangle$ and ${\cal{L}}$ is completely positive, it holds that
$$
\nu_\psi(n) \geq 0,\;\;\forall |\psi\rangle,\forall n.
$$
We may now evaluate $\nu_\psi(n)$ in a different way. Using Eq.~(\ref{nu_def}), in the limit of $n\rightarrow \infty$ we have
$$
\begin{array}{rcl}
\nu_\psi(\infty) &=&  \langle \psi | \text{Tr}_\text{junk}\left(\left( \sum_{ij} |i\rangle \nu_{ij} \langle j| \right)  \otimes \rho_\text{fix} \right)|\psi\rangle\\
&=& \text{Tr}(\rho_\text{fix}) \times \langle \psi | \left(\sum_{ij} |i\rangle \nu_{ij} \langle j|\right) |\psi\rangle\\
&=& \text{Tr}(\rho_\text{fix}) \times \langle \psi |  \nu  |\psi\rangle.
\end{array}
$$
Comparing the two above expressions for $\nu_\psi(\infty)$, we find that the state $\nu$ is positive semidefinite, as claimed.\medskip

\begin{figure}
\begin{center}
\includegraphics[width=8cm]{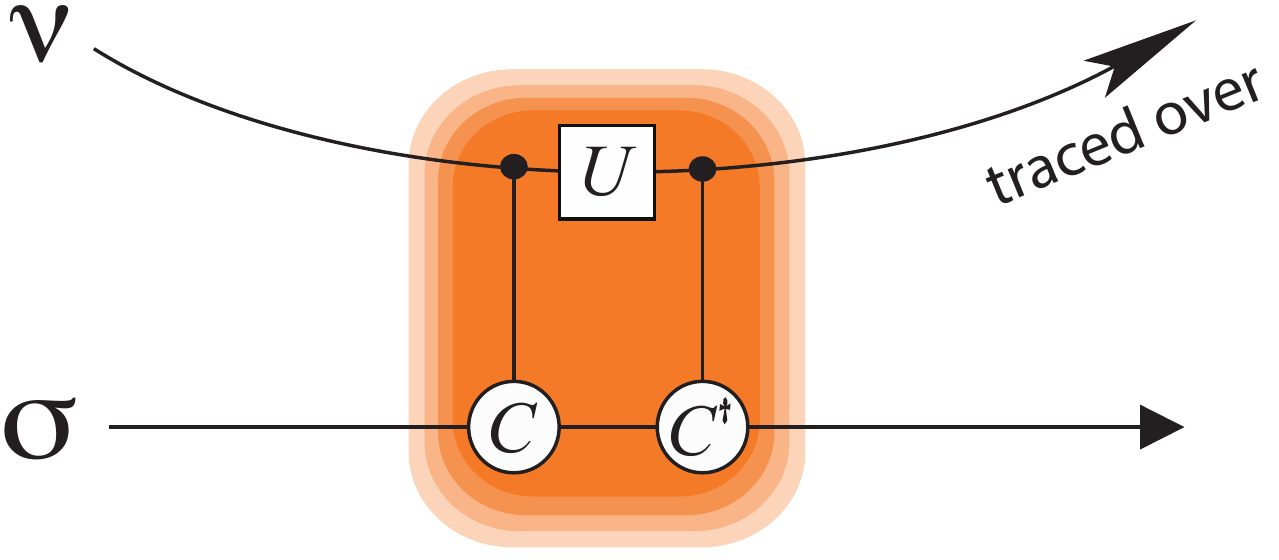}
\caption{\label{Int}Computation by interaction of the logical system with the state $\nu$. The state $\nu$ represents the ``computational essence'' of the fixed point state $\rho_\text{fix}$ of the junk subsystem. $U$ is the unitary that transforms between the wire basis and the measurement basis for the physical degree of freedom. The interaction between the two systems is via the gate Eq.~(\ref{cC}).}
\end{center}
\end{figure}

Now that we know that $\nu$ can be regarded as a quantum state, what is its role in the computational scheme? It turns out that each copy of the state $\nu$ implements a single quantum gate on the logical subsystem, by interacting with it and subsequently being discarded (traced out). See Fig.~\ref{Int} for illustration. 

This picture arises as follows. First, consider the action on the virtual state $\sigma \otimes \rho_\text{fix}$ of a physical degree of freedom being measured in the state $|\psi_0\rangle$ (corresponding to byproduct operator $I$) followed by oblivious wire ${\cal{L}}^n$ for some sufficiently large $n$. With Eq.~(\ref{nu_def}),
$$
\begin{array}{rcl}
\sigma\otimes \rho_\text{fix} &\longrightarrow& {\cal{L}}^n\left( \left[\sum_{i}\langle \psi_0|i\rangle C_i\otimes B_i \right]\left(\sigma\otimes \rho_\text{fix}\right) \right)\\
&=&\sum_{ij} \langle  \psi_0| i\rangle \nu_{ij}   \langle j |\psi_0\rangle C_i  \sigma C_j^\dagger  \otimes \rho_\text{fix}.
\end{array}
$$
Now we consider the following related procedure, which invokes the incoherent addition of computational paths. The physical degree of freedom is measurement in the basis ${\cal{B}}_M=\{|\psi_k\rangle, k=0,..,d-1\}$ which is related to the wire basis ${\cal{B}}$ via a unitary $U$, $|\psi_k \rangle = U |k\rangle$, for all $|k\rangle \in {\cal{B}}$. For each measurement outcome $k$, the corresponding byproduct operator $C_k$ is reversed; See Section~\ref{iG}. After that, the measurement outcome $k$ is discarded. Combining all computational paths $k$, this procedure is represented by
$$
\sigma \longrightarrow \sum_k  \sum_{ij} \langle  k| U^\dagger | i\rangle \nu_{ij}   \langle j | U |k\rangle C_k^\dagger C_i \sigma C_j^\dagger C_k.
$$
Here, we have dropped the tensor factor $\rho_\text{fix}$ since it remains unaffected by the procedure. We note that a new tensor factor comes into play, namely the physical degree of freedom, with states $\{|i\rangle\}$. We denote by $\Lambda(C)$ the entangling gate
\begin{equation}\label{cC}
\Lambda(C)=\sum_{i=0}^{d-1} |i\rangle \langle i| \otimes C_i,
\end{equation}
with the physical degree of freedom as control and the logical subsystem as target. With this, the update of the state $\sigma$ can be rewritten as
$$
\sigma \longrightarrow \text{Tr}_{\cal{P}}  \Lambda(C)^\dagger  U \Lambda(C)   \left( \nu \otimes  \sigma \right)  \Lambda(C)^\dagger U^ \dagger \Lambda(C),
$$
where $\text{Tr}_{\cal{P}}$ denotes the trace over the physical degree of freedom. This is the state evolution depicted in Fig.~\ref{Int}. We have thus shown that MBQCs in SPTO phases simulate quantum circuits in which each gate be viewed as arising from the interaction between the logical subsystem with a particle prepared in the state $\nu$.

\section{Conclusion}

We have investigated measurement-based quantum computation on resource states that are  ground states in symmetry-protected topologically ordered phases, in spatial dimension one. We have shown that if an SPTO phase supports symmetry-protected quantum wire (and subject to the technical condition that the MBQC byproduct operators are {elements of} a projective representation of the symmetry group), then this phase also supports quantum computation on one qudit. That is, any state in the phase is a resource for MBQC in ${\cal{D}}=1$. {For such SPTO phases, Conjecture~\ref{MC} thus holds.}

It is instructive to compare the present construction to the symmetry-protected wire of \cite{Else}, which it extends to the domain of quantum computation. In \cite{Else}, symmetry specifies the computational wire completely, and the details of the quantum state in the given SPTO phase play no role at all. This is different in the present scenario for quantum computation. Some information about the resource quantum state does enter the computational scheme, namely a Hermitian $d\times d$-matrix of coefficients $[\nu_{ij}]$, with $d$ the dimension of the physical spins in the chain. These parameters are not constrained by symmetry, and need to be measured in a self-test prior to the computation.   

We conclude with two open questions. The immediate question is whether and how Theorem~\ref{MT} generalizes to higher spatial dimension ${\cal{D}}\geq 2$.

Even in dimension ${\cal{D}}=1$, a better understanding of the algebraic side of the presented constructions is desirable. For example, can it happen that a byproduct operator $C_i\otimes I$ is not a  symmetry? Furthermore, is the existence of a wire basis necessary for uniform computational power throughout SPTO phases? We have found that the MPS tensors are often very constrained by symmetry, but the constraints are not quite strong enough or not of the right kind to produce a tensor product structure between a logical and a junk system. Can computational schemes that work uniformly across SPTO phases be built on other structures than tensor products?

\acknowledgments{This work is supported by NSERC, Cifar, and the National Science Foundation under Grant No. PHY 1620252. R. Raussendorf is fellow of the Cifar program on Quantum Information Science.}

\appendix

\section{Adding computational paths---the one-party scenario}\label{1P}

Here we return to the subject of Section~\ref{F/D}, the interpretation of adding computational paths. We have already provided an interpretation for a distributed computational setting involving two parties. However, in the present situation, the computation is performed by a single party. In this setting, we use ``forgetting'' of classical information in the following technical sense: Forgetting of information means the commitment to never use that information in any decision-making after the point of ``forgetting''. It is thus merely a restriction in the available modes of classical processing, and we do not have to discuss whether the erasure of information is physically possible.

A general quantum computation consists of an initialization, followed by a sequence  of unitaries and measurements. Since we have not established all of those computational primitives yet, we consider here the simple case where oblivious wire is implemented on a state $|\tilde{\Phi}\rangle$ with a left boundary state $|L\rangle$, and subsequently an observable $A$ is measured on the logical subsystem of the right boundary state. Furthermore, the left boundary state is of the special tensor product form $|L\rangle=|l\rangle \otimes |j\rangle$, where $|l\rangle$ represents the state of the logical subsystem of the left boundary, and $|j\rangle$ the state of the junk subsystem.

{The important quantity in this setting is the probability distribution $p_A$ of measurement outcomes for $A$ given the state $|l\rangle$. The goal is to show that the distribution $p_A$ can be sampled from correctly and efficiently by measurements on $|\tilde{\Phi}\rangle$.}

{We denote by $q_A(o,\textbf{s})$ the joint probability for obtaining the measurement record $\textbf{s}$ on the spins in the chain and the outcome $o$ for the observable $A$, measured  on the right boundary system in the support of $|\tilde{\Phi}\rangle$. By the same argument as in Eqs.~(\ref{taur}), (\ref{Taur}), but without summation over $\textbf{s}$, we find that
\begin{equation}\label{qqp}
q_A(\textbf{s},o) = q(\textbf{s})\,p_A(o|\textbf{s}),
\end{equation}
where $q(\textbf{s})=|\kappa(n)|^2\|\prod_{k=1}^n B_{s_k} |j\rangle\|^2$ is the probability of obtaining the outcome $\textbf{s}$, and 
$$
p_A(o|\textbf{s})=\langle l| P_A(o)|l\rangle
$$ 
is the probability of obtaining the outcome $o$ in the measurement of the right boundary system, given the prior measurement record $\textbf{s}$. Therein, $P_A(o)$ is the projector onto the eigenspace of $A$ with eigenvalue $o$.}

{We thus find that $p_A(o|\textbf{s})$ is independent of $\textbf{s}$, and, more importantly, that it equals the probability of obtaining the outcome $o$ in the measurement of the observable $A$ on the state $|l\rangle$, $p_A(o|\textbf{s})=p_A(o)$ for all $\textbf{s}$.  Thus, with Eq.~(\ref{qqp}),
\begin{equation}\label{pAq}
p_A(o)= \sum_\textbf{s}q_A(\textbf{s},o).
\end{equation}
We now turn to the experimental procedure.
In each run of the computation, the computing party obtains one sample $(\textbf{s},o)$ from the the probability distribution $q_A$, and discards the $\textbf{s}$-part,
$$
(o,\textbf{s}) \longrightarrow o.
$$
This is the step of ``forgetting'' classical information. The remaining outcomes $o$ are thus sampled from the probability distribution $p'_A$, with
$$
p'_A(o):=\sum_\textbf{s} q_A(\textbf{s},o).
$$
By comparison with Eq.~(\ref{pAq}), $p'_A\equiv p_A$. The procedure thus samples from the correct probability distribution for the measurement outcomes $o$ of $A$ on $|l\rangle$.}

{Furthermore, every run of the computation generates a sample from the distribution $p_A$, and the adding of computational paths is thus efficient. Specifically, it does not cause any overhead in the computation.}

\section{Proof of Lemma~\ref{Concat}}\label{conc}

{\em{Proof of Lemma~\ref{Concat}.}} The proof is by explicit construction of the procedure of implementing $T'(\beta) T(\alpha)$. This procedure is the following: (i) On the first segment of the spin chain, implement $T(\alpha)$, but retain the measurement outcomes $(s_1,.., s_n)=\textbf{s}$. (ii) On the remaining segment of the spin chain, implement $T'(\beta'(\textbf{s}))$. (iii) Sum over all measurement outcomes $\textbf{s}$ on the first segment and $\textbf{z}=(z_{n+1},..,z_{n'})$ on the second segment.

We denote by $[a]$ the super operator corresponding to the operator $a$, and $C(\textbf{z},\textbf{s})$ is the total accumulated byproduct operator. The output state $\tau_\text{out}$ of the combined procedure is
\begin{widetext}
$$
\begin{array}{rcl}
\tau_\text{out} &=& \displaystyle{\sum_\textbf{z}[C(\textbf{z},\textbf{s})]^{-1}  \prod_{j=n+2}^{n'} [C_{z_j}\otimes B_{z_j}] [A_{n+1}(z_{n+1},\beta'(\textbf{s}))] \sum_\textbf{s} \prod_{i=2}^{n} [C_{s_i}\otimes B_{s_i}] [A_1(s_1,\alpha)] \big( \sigma \otimes \rho_\text{fix} \big)} \\
&=& \displaystyle{\sum_\textbf{z}[C(\textbf{z},\textbf{s})]^{-1} \left[C(\textbf{z})\otimes \prod_{j=n+2}^{n'} B_{z_j}\right] [A_{n+1}(z_{n+1},\beta'(\textbf{s}))] \sum_\textbf{s} \left[C(\textbf{s})\otimes \prod_{i=2}^{n} B_{s_i}\right] [A_1(s_1,\alpha)] \big( \sigma \otimes \rho_\text{fix} \big)} \\
&=& \displaystyle{\sum_{\textbf{z},\textbf{s}}[C(\textbf{z} ,\textbf{s})]^{-1} \left[C(\textbf{z}) C(\textbf{s})\otimes \prod_{j=n+2}^{n'} B_{z_j}\right] [A_{n+1}(z_{n+1},\beta)] \left[I\otimes \prod_{i=2}^{n} B_{s_i}\right] [A_1(s_1,\alpha)] \big( \sigma \otimes \rho_\text{fix} \big)} \\
&=& \displaystyle{\sum_{\textbf{z},\textbf{s}}  \left[\prod_{j=n+2}^{n'} I \otimes B_{z_j}\right] [A_{n+1}(z_{n+1},\beta)] \left[\prod_{i=2}^{n}I \otimes  B_{s_i}\right] [A_1(s_1,\alpha)] \big( \sigma \otimes \rho_\text{fix} \big)} \\
&=& \displaystyle{\sum_\textbf{z}  \left[\prod_{j=n+2}^{n'} I \otimes B_{z_j}\right] [A_{n+1}(z_{n+1},\beta)] \sum_\textbf{s}\left[\prod_{i=2}^{n}I \otimes  B_{s_i}\right] [A_1(s_1,\alpha)] \big( \sigma \otimes \rho_\text{fix} \big)} \\
&=& \displaystyle{\sum_\textbf{z}  \left[\prod_{j=n+2}^{n'} I \otimes B_{z_j}\right] [A_{n+1}(z_{n+1},\beta)] T(\alpha)  \big( \sigma  \big) \otimes \rho_\text{fix}} \\
&=& \displaystyle{T'(\beta)T(\alpha) \big( \sigma \big) \otimes \rho_\text{fix}} \\
\end{array}
$$
\end{widetext}
Therein, in the third line we have used Eq.~(\ref{BasFlip}), and in the fourth line the relation $C(\textbf{s},\textbf{z})=C(\textbf{z})C(\textbf{s})$. $\Box$


\begin{thebibliography}{99}

\bibitem{RB01}
R. Raussendorf and H.-J. Briegel, Phys. Rev. Lett. \textbf{86}, 5188 (2001).

\bibitem{BR}
H.-J. Briegel and R. Raussendorf, Phys. Rev. Lett. \textbf{86}, 910 (2001).

\bibitem{AKLT}
I. Affleck, T. Kennedy, E. H. Lieb, and H. Tasaki, Phys. Rev. Lett. \textbf{59}, 799 (1987).

\bibitem{Aka0}
A. Miyake, Annals of Physics \textbf{326}, 1656 (2011).

\bibitem{WAR}
T.-C. Wei, I. Affleck and R. Raussendorf, Phys. Rev. Lett \textbf{106}, 070501 (2011).

\bibitem{GFE}
D. Gross, S.T. Flammia, J. Eisert, Phys. Rev. Lett. 1\textbf{102},190501 (2009).

\bibitem{BMW}
M.J. Bremner, C. Mora, A Winter, Phys. Rev. Lett. \textbf{102}, 190502 (2009).

\bibitem{SPT1}
Xie Chen, Zheng Cheng Gu, and Xiao Gang Wen, Phys. Rev. B \textbf{83}, 035107 (2011).

\bibitem{SPT2}
Xie Chen, Zheng-Cheng Gu, Zheng-Xin Liu, and Xiao-Gang Wen, Phys. Rev. B \textbf{87}, 155114 (2013).

\bibitem{SPT3}
Norbert Schuch, David Perez-Garcia, and Ignacio Cirac,  Phys. Rev. B  \textbf{84}, 165139 (2011).

\bibitem{Aka1}
Akimasa Miyake,  Phys. Rev. Lett. \textbf{105},1 (2010).

\bibitem{Else}
D.V. Else, I. Schwarz, S.D. Bartlett and A. Doherty, Phys. Rev. Lett. \textbf{108}, 240505 (2012).

\bibitem{Aka3}
Jacob Miller and Akimasa Miyake, Phys. Rev. Lett. 114, 120506 (2015).

\bibitem{PraWei}
A. Prakash and T.-C. Wei, Phys. Rev. A \textbf{92}, 022310 (2015).

\bibitem{PraWei2}
A. Prakash, C. G. West, and T.-C. Wei, arXiv: 1604.00037.

\bibitem{WR2}
Tzu-Chieh Wei, Poya Haghnegahdar, Robert Raussendorf,   Phys. Rev. A \textbf{90}, 042333 (2014).

\bibitem{WR3}
Tzu-Chieh Wei, Robert Raussendorf, Phys. Rev. A \textbf{92}, 012310 (2015).

\bibitem{Zitartz}
H. Niggemann, A. Kl{\"u}mper, and J. Zittartz, Z. Phys. B \textbf{104}, 103 (1997).

\bibitem{Darmawan}
Andrew S. Darmawan, Gavin K. Brennen, Stephen D. Bartlett,  New J. Phys. 14, 013023 (2012).

\bibitem{HWW}
Ching-Yu Huang, Maximilian Anton Wagner, and Tzu-Chieh Wei, arXiv:1605.08417.

\bibitem{MPS}
D. Perez-Garcia, F. Verstraete, M.M. Wolf, J.I. Cirac, Quantum Inf. Comput. \textbf{7}, 401 (2007).

\bibitem{Leung}
Andrew M. Childs, Debbie W. Leung, Michael A. Nielsen, Phys. Rev. A \textbf{71}, 032318 (2005).

\bibitem{Eisert}
D. Gross, J. Eisert, Phys. Rev. Lett. \textbf{98}, 220503 (2007).

\bibitem{QCcoho}
D. Stephen, Dong-Sheng Wang, Abhishodh Prakash, Tzu-Chieh Wei and Robert Raussendorf, in preparation.

\bibitem{QETP}
Bei Zeng, Xie Chen, Duan-Lu Zhou, Xiao-Gang Wen, arXiv:1508.02595.

\end{thebibliography}
\end{document}